
\input harvmac.tex
\input epsf

\def\defn#1{\bigskip\noindent{\bf Definition #1} }

\def\rmk#1{\bigskip\noindent{\bf Remarks} }


\def\unlockat{\catcode`\@=11}
\def\lockat{\catcode`\@=12}

\unlockat

\def\newsec#1{\global\advance\secno by1\message{(\the\secno. #1)}
\global\subsecno=0\global\subsubsecno=0\eqnres@t\noindent
{\bf\the\secno. #1}
\writetoca{{\secsym} {#1}}\par\nobreak\medskip\nobreak}
\global\newcount\subsecno \global\subsecno=0
\def\subsec#1{\global\advance\subsecno
by1\message{(\secsym\the\subsecno. #1)}
\ifnum\lastpenalty>9000\else\bigbreak\fi\global\subsubsecno=0
\noindent{\it\secsym\the\subsecno. #1}
\writetoca{\string\quad {\secsym\the\subsecno.} {#1}}
\par\nobreak\medskip\nobreak}
\global\newcount\subsubsecno \global\subsubsecno=0
\def\subsubsec#1{\global\advance\subsubsecno by1
\message{(\secsym\the\subsecno.\the\subsubsecno. #1)}
\ifnum\lastpenalty>9000\else\bigbreak\fi
\noindent\quad{\secsym\the\subsecno.\the\subsubsecno.}{#1}
\writetoca{\string\qquad{\secsym\the\subsecno.\the\subsubsecno.}{#1}}
\par\nobreak\medskip\nobreak}

\def\subsubseclab#1{\DefWarn#1\xdef
#1{\noexpand\hyperref{}{subsubsection}%
{\secsym\the\subsecno.\the\subsubsecno}%
{\secsym\the\subsecno.\the\subsubsecno}}%
\writedef{#1\leftbracket#1}\wrlabeL{#1=#1}}
\lockat


\def\figin{\epsfcheck\figin}\def\figins{\epsfcheck\figins}
\def\epsfcheck{\ifx\epsfbox\UnDeFiNeD
\message{(NO epsf.tex, FIGURES WILL BE IGNORED)}
\gdef\figin##1{\vskip2in}\gdef\figins##1{\hskip.5in}
\else\message{(FIGURES WILL BE INCLUDED)}%
\gdef\figin##1{##1}\gdef\figins##1{##1}\fi}
\def\DefWarn#1{}
\def\figinsert{\goodbreak\midinsert}
\def\ifig#1#2#3{\DefWarn#1\xdef#1{fig.~\the\figno}
\writedef{#1\leftbracket fig.\noexpand~\the\figno}%
\figinsert\figin{\centerline{#3}}\medskip\centerline{\vbox{\baselineskip12pt
\advance\hsize by -1truein\noindent\footnotefont{\bf Fig.~\the\figno:} #2}}
\bigskip\endinsert\global\advance\figno by1}


\def\boxit#1{\vbox{\hrule\hbox{\vrule\kern8pt
\vbox{\hbox{\kern8pt}\hbox{\vbox{#1}}\hbox{\kern8pt}}
\kern8pt\vrule}\hrule}}
\def\mathboxit#1{\vbox{\hrule\hbox{\vrule\kern8pt\vbox{\kern8pt
\hbox{$\displaystyle #1$}\kern8pt}\kern8pt\vrule}\hrule}}

\def\IL{\relax{\rm I\kern-.18em L}}
\def\IH{\relax{\rm I\kern-.18em H}}
\def\IR{\relax{\rm I\kern-.18em R}}
\def\IC{\relax\hbox{$\inbar\kern-.3em{\rm C}$}}
\def\IZ{\relax\ifmmode\mathchoice
{\hbox{\cmss Z\kern-.4em Z}}{\hbox{\cmss Z\kern-.4em Z}}
{\lower.9pt\hbox{\cmsss Z\kern-.4em Z}}
{\lower1.2pt\hbox{\cmsss Z\kern-.4em Z}}\else{\cmss Z\kern-.4em
Z}\fi}

\def\CN {{\cal N}}
\def\CR {{\cal R}}
\def\CD {{\cal D}}
\def\CF {{\cal F}}

\def\CP {{\cal P }}

\def\CO {{\cal O}}
\def\CZ {{\cal Z}}

\def\CC {{\cal C}}

\def\CK{{\cal K}}


\def\CN {{\cal N}}

\def\CO {{\cal O}}

\def\CP {{\cal P }}

\def\CZ {{\cal Z }}

\font\manual=manfnt \def\dbend{\lower3.5pt\hbox{\manual\char127}}

\def\IZ{\relax\ifmmode\mathchoice
{\hbox{\cmss Z\kern-.4em Z}}{\hbox{\cmss Z\kern-.4em Z}}
{\lower.9pt\hbox{\cmsss Z\kern-.4em Z}}
{\lower1.2pt\hbox{\cmsss Z\kern-.4em Z}}\else{\cmss Z\kern-.4em
Z}\fi}
\def\half {{1\over 2}}

\def\p{\partial}

\def\CN {{\cal N}}

\def\CO {{\cal O}}

\def\CP {{\cal P }}

\def\CZ {{\cal Z }}


\def\IZ{\relax\ifmmode\mathchoice
{\hbox{\cmss Z\kern-.4em Z}}{\hbox{\cmss Z\kern-.4em Z}}
{\lower.9pt\hbox{\cmsss Z\kern-.4em Z}}
{\lower1.2pt\hbox{\cmsss Z\kern-.4em Z}}\else{\cmss Z\kern-.4em
Z}\fi}
\def\IB{\relax{\rm I\kern-.18em B}}
\def\IC{{\relax\hbox{$\inbar\kern-.3em{\rm C}$}}}
\def\ID{\relax{\rm I\kern-.18em D}}
\def\IE{\relax{\rm I\kern-.18em E}}
\def\IF{\relax{\rm I\kern-.18em F}}
\def\IG{\relax\hbox{$\inbar\kern-.3em{\rm G}$}}
\def\IGa{\relax\hbox{${\rm I}\kern-.18em\Gamma$}}
\def\IH{\relax{\rm I\kern-.18em H}}
\def\II{\relax{\rm I\kern-.18em I}}
\def\IK{\relax{\rm I\kern-.18em K}}
\def\IP{\relax{\rm I\kern-.18em P}}

\def\IQ{\relax\hbox{$\inbar\kern-.3em{\rm Q}$}}
\def\IP{\relax{\rm I\kern-.18em P}}

\def\im{{\rm Im}}

\def\inbar{\,\vrule height1.5ex width.4pt depth0pt}

\def\mod{{\rm mod}}
\def\p{\partial}

\font\cmss=cmss10 \font\cmsss=cmss10 at 7pt
\def\IR{\relax{\rm I\kern-.18em R}}

\def\Tr{{\rm Tr}}


\def\boxit#1{\vbox{\hrule\hbox{\vrule\kern8pt
\vbox{\hbox{\kern8pt}\hbox{\vbox{#1}}\hbox{\kern8pt}}
\kern8pt\vrule}\hrule}}
\def\mathboxit#1{\vbox{\hrule\hbox{\vrule\kern8pt\vbox{\kern8pt
\hbox{$\displaystyle #1$}\kern8pt}\kern8pt\vrule}\hrule}}


\def\inbar{\,\vrule height1.5ex width.4pt depth0pt}

\def\p{\partial}

\font\cmss=cmss10 \font\cmsss=cmss10 at 7pt
\def\IR{\relax{\rm I\kern-.18em R}}

\def\Tr{\rm Tr}


\def\makeblankbox#1#2{\hbox{\lower\dp0\vbox{\hidehrule{#1}{#2}%
   \kern -#1
   \hbox to \wd0{\hidevrule{#1}{#2}%
      \raise\ht0\vbox to #1{}
      \lower\dp0\vtop to #1{}
      \hfil\hidevrule{#2}{#1}}%
   \kern-#1\hidehrule{#2}{#1}}}%
}%
\def\hidehrule#1#2{\kern-#1\hrule height#1 depth#2 \kern-#2}%
\def\hidevrule#1#2{\kern-#1{\dimen0=#1\advance\dimen0 by #2\vrule
    width\dimen0}\kern-#2}%
\def\openbox{\ht0=1.2mm \dp0=1.2mm \wd0=2.4mm  \raise 2.75pt
\makeblankbox {.25pt} {.25pt}  }
\def\opensquare{\ht0=3.4mm \dp0=3.4mm \wd0=6.8mm  \raise 2.7pt \makeblankbox
{.25pt} {.25pt}  }

\def\sector#1#2{\ {\scriptstyle #1}\hskip 1mm
\mathop{\opensquare}\limits_{\lower 1mm\hbox{$\scriptstyle#2$}}\hskip 1mm}

\def\tsector#1#2{\ {\scriptstyle #1}\hskip 1mm
\mathop{\opensquare}\limits_{\lower 1mm\hbox{$\scriptstyle#2$}}^\sim\hskip 1mm}

\def\holographic{ fareytail }
\def\Holographic{ Fareytail }


\lref\apostol{T. Apostol, {\it Modular Functions and
Dirichlet Series in Number Theory}, Springer Verlag 1990}
\lref\breckenridge{J.C. Breckenridge,  D.A. Lowe,  R.C. Myers,  A.W. Peet,
A. Strominger, C. Vafa,
``Macroscopic and Microscopic Entropy of Near-Extremal Spinning Black Holes,''
hep-th/9603078; Phys.Lett. B381 (1996) 423-426}
 \lref\carlipi{S. Carlip, ``The $(2+1)$-Dimensional
Black Hole,'' gr-qc/9506079}
\lref\carlipii{S. Carlip and C. Teitelboim}
\lref\cveticyoum{M. Cvetic and D. Youm,
``General Rotating Five Dimensional Black Holes of Toroidally Compactified
Heterotic String,''
hep-th/9603100; Nucl.Phys. B476 (1996) 118-132}
\lref\ez{M. Eichler and D. Zagier, {\it The theory of
Jacobi forms}, Birkh\"auser 1985}
\lref\emss{S. Elitzur, G. Moore,
A. Schwimmer, and N. Seiberg, ``Remarks on the Canonical Quantization of
the Chern-Simons-Witten Theory,''
Nucl. Phys. {\bf B326}(1989)108}
\lref\tkawai{T. Kawai, ``K3 surfaces, Igusa cusp form and
string theory,'' hep-th/9710016}
\lref\kawai{T. Kawai, Y. Yamada, and S. -K. Yang,
``Elliptic genera and $N=2$ superconformal
field theory,'' hepth/9306096 }
\lref\maldacenai{J. Maldacena, ``Black holes and
D-branes,'' hep-th/9705078}
\lref\greybody{J. Maldacena and A. Strominger, Greybody
factors}
\lref\stringexclusion{J. Maldacena and A. Strominger,
``$AdS_3$ black holes and a stringy exclusion
principle,'' hep-th/9804085}
\lref\rademacheri{H. Rademacher, Topics in
Analytic Number Theory}
\lref\rademacherii{H. Rademacher, {\it Lectures on
Elementary Number Theory}, Robert E. Krieger Publishing
Co. , 1964}
\lref\seibschw{N. Seiberg and A. Schwimmer,
``Comments on the N=2,N=3,N=4 superconformal
algebras in two dimensions,'' Phys. Lett.
{\bf 184B}(1987)191 }
\lref\strdec{A. Strominger, December paper}
\lref\townsend{A. Achucarro and P.K. Townsend,
Phys. Lett. {\bf 180B}(1986) 89; J.M. Izquierdo
and P.K. Townsend, ``Supersymmetric spacetimes in
$(2+1)$ adS-supergravity models,'' gr-qc/9501018}
\lref\warner{N. Warner, ``Lectures on $N=2$ superconformal
theories and singularity theory,'' in Superstrings 89}
\lref\wittenads{E. Witten, ``Anti- de Sitter Space and
holography,'' hep-th/9802150; ``Anti-de Sitter space,
thermal phase transition, and confinement in
gauge theories,'' hep-th/9803131}

\lref\vafawitten{C. Vafa and E. Witten, ``A Strong Coupling Test
of S-Duality,'' hep-th/9804074, Nucl.Phys. B431 (1994) 3-77.}


\lref\btz{M.  Ba\~ nados. C.   Teitelboim, and J. Zanelli,
``The Black Hole in Three Dimensional Space Time,''
hep-th/9204099; Phys.Rev.Lett. 69 (1992) 1849-1851}

\lref\bateman{Bateman Manuscript project}

\lref\bost{L. Alvarez-Gaum\'e, J.-B. Bost, G. Moore,
P. Nelson, and C. Vafa, ``Bosonization on higher
genus Riemann surfaces,'' Commun.Math.Phys.112:503,1987 }

\lref\cardi{ K. Behrndt, G. Lopes Cardoso, B. de Wit, R. Kallosh, D. Lvst, T.
Mohaupt, ``Classical and
quantum N=2 supersymmetric black holes,''
hep-th/9610105}

\lref\bloch{S. Bloch, ``The proof of the Mordell
conjecture,'' Math. Intell. {\bf 6}(1984) 41}

\lref\borcherds{R.E. Borcherds, ``Reflection groups of Lorentzian
lattices,'' math.GR/9909123}
 \lref\cassels{J.W.S. Cassels, {\it
Lectures on Elliptic Curves}, Cambridge University Press, 1995}

\lref\cox{D.A. Cox, {\it Primes of the form $x^2 + n y^2$}, John Wiley, 1989.}

\lref\cremona{J.E. Cremona, {\it Algorithms for modular
elliptic curves} Cambridge University Press 1992}
\lref\davenport{H. Davenport, {\it Multiplicative
Number Theory}, Second edition, Springer-Verlag,
GTM 74}

\lref\dewit{Bernard de Wit, Gabriel Lopes Cardoso, Dieter L\"ust,
 Thomas Mohaupt, Soo-Jong Rey,
``Higher-Order Gravitational Couplings and Modular Forms in $N=2,D=4$ Heterotic
String Compactifications,''
hep-th/9607184; G.L. Cardoso, B. de Wit, and
T. Mohaupt, ``Corrections to macroscopic supersymmetric
black hole entropy,'' hep-th/9812082}
\lref\birmingham{D. Birmingham, C. Kennedy,
S. Sen, and A. Wilkins, ``Geometrical finiteness,
holography, and the BTZ black hole,''
hep-th/9812206}

\lref\dvv{R. Dijkgraaf, E. Verlinde, and H. Verlinde,
``Counting Dyons in N=4 String Theory,''
hep-th/9607026;Nucl.Phys. B484 (1997) 543-561}

\lref\duke{W. Duke, ``Hyperbolic distribution problems
and half-integral weight Maass forms,'' Invent. Math.
{\bf 92} (1988) 73}
\lref\faltings{G. Faltings, ``Finiteness theorems for abelian
varieties over number fields,'' in {\it Arithmetic Geometry},
G. Cornell and J.H. Silverman, eds. Springer 1986}
\lref\fks{S. Ferrara, R. Kallosh, and A. Strominger,
``N=2 Extremal Black Holes,''   hep-th/9508072}
\lref\fk{S. Ferrara and R. Kallosh, ``Universality of Sypersymmetric
Attractors,''   hep-th/9603090;  ``Supersymmetry and Attractors,''
hep-th/9602136; S. Ferrara, ``Bertotti-Robinson Geometry and Supersymmetry,''
hep-th/9701163}
\lref\enneightbh{ L. Andrianopoli, R. D'Auria,
S. Ferrara, P. Fre', M. Trigiante,
`` $E_{7(7)}$ Duality, BPS Black-Hole Evolution and Fixed Scalars,''
 hep-th/9707087; Nucl.Phys. B509 (1998) 463-518
 }

\lref\fm{S. Ferrara and J. Maldacena, ``Branes, central
charges and $U$-duality invariant BPS conditions,''
hep-th/9706097}

\lref\husemoller{D. Husemoller, {\it Elliptic Curves},
Springer-Verlag, GTM 111}

\lref\faltings{G. Faltings, ``Calculus on arithmetic
surfaces,'' Ann. Math. {\bf 119}(1984) 387}

\lref\bgross{B. Gross, {\it Arithmetic on elliptic
curves with complex multiplication}, Springer-Verlag LNM  776}

\lref\gzsm{B. Gross and D. Zagier,
``On singular moduli,'' J. reine angew. Math.
{\bf 355} (1985) 191}

\lref\kalkol{R. Kallosh and B. Kol,
``E(7) Symmetric Area of the Black Hole Horizon,''
hep-th/9602014}

\lref\knopp{M.I. Knopp, ``Rademacher on $J(\tau)$,
Poincar\'e series of nonpositive weights and the
Eichler cohomology,'' Notices of the Amer. Math.
Soc. {\bf 37}(1990) 385}

\lref\msw{J. Maldacena, A. Strominger, and E. Witten,
``Black Hole Entropy in M-Theory,''  hep-th/9711053}

\lref\arthatt{G. Moore, ``Arithmetic and Attractors,''
hep-th/9807087 ; ``Attractors and Arithmetic,'' hep-th/9807056}

\lref\fareytail{ ``A Black Hole Farey Tail,''
unpublished. To appear, some day, somewhere, maybe,....}

\lref\lang{S. Lang, {\it Introduction to Arakelov
Geometry}, Springer-Verlag 1988}

\lref\langalg{S. Lang, {\it Algebra}, Addison-Wesley 1970}

\lref\langelliptic{S. Lang, {\it Elliptic Functions},
Addison-Wesley, 1973}

\lref\littlewood{J.E. Littlewood, ``On the class number of
the corpus $P(\sqrt{-k})$,'' Proc. Lond. Math. Soc., Ser. 2, {\bf 27},
Part 5. (1928)}

\lref\mazur{B. Mazur, ``Arithmetic on curves,'' Bull.
Amer. Math. Soc. {\bf 14}(1986) 207}

\lref\miller{S.D. Miller, in preparation.}

\lref\millermoore{S.D. Miller and G. Moore, ``Landau-Siegel Zeroes
and Black Hole Entropy,'' hep-th/9903267}

 \lref\mumford{D. Mumford, Tata Lectures on
Theta}

\lref\peet{A. Peet, ``The Bekenstein Formula and String Theory (N-brane
Theory),''
hep-th/9712253}

\lref\rohrlich{D. Rohrlich, ``Elliptic curves with
good reduction everywhere,'' J. London Math. Soc.,
{\bf 25}(1982)216}

\lref\silvermanag{J. Silverman, {\it Arithmetic Geometry},
G. Cornell and J.H. Silverman, eds. Springer Verlag 1986}

\lref\silveraec{J. Silverman, {\it The Arithmetic of Elliptic
Curves}, Springer Verlag GTM 106 1986}

\lref\silveradvtop{J. Silverman, {\it Advanced Topics in the
Arithmetic of Elliptic Curves} Springer Verlag GTM 151, 1994}

\lref\stromi{A. Strominger, ``Macroscopic Entropy of $N=2$ Extremal Black
Holes,''  hep-th/9602111}

\lref\sv{A. Strominger and C. Vafa,
``Microscopic Origin of the Bekenstein-Hawking Entropy,''
hep-th/9601029; Phys.Lett. B379 (1996) 99-104 }

\lref\tatuzawa{T. Tatuzawa, ``On a theorem of
Siegel,'' Jap. J. Math. {\bf 21}(1951)163 }

\lref\weil{A. Weil, {\it Elliptic functions according to
Eisenstein and Kronecker}, Springer-Verlag 1976}

\lref\wittenls{E. Witten, `` Phases of $N=2$ Theories In Two Dimensions,''
hep-th/9301042;Nucl.Phys. B403 (1993) 159-222}

\lref\review{  O. Aharony, S.S. Gubser, J. Maldacena, H. Ooguri, Y. Oz,
``Large N Field Theories, String Theory and Gravity,'' hep-th/9905111. }
\lref\cvl{M. Cveti\'c and F. Larsen, ``Near Horizon Geometry
of Rotating Black Holes in Five Dimensions,''
hep-th/9805097}
\lref\martsahak{E. Martinec and V. Sahakian,
``Black holes and five-brane thermodynamics,''
hep-th/9901135; Phys. Rev. {\bf D60}(1999)
064002}
\lref\carlipsum{S. Carlip,
``The sum over topologies in three-dimensional
Euclidean quantum gravity,''
hep-th/9206103; Class.Quant.Grav.10:207-218,1993.
S. Carlip, ``Dominant topologies in
Euclidean quantum gravity,''
gr-qc/9710114; Class.Quant.Grav.15:2629-2638,1998.
 }
\lref\degersezgin{S. Deger, A. Kaya, E. Sezgin, and P. Sundell,
``Spectrum of D=6, N=4b supergravity on $AdS_3 \times S^3$,''
hep-th/9804166}
 \lref\hardyram{G.H. Hardy and S. Ramanujan, ``Asymptotic
formulae in combinatory analysis,'' Proc. Lond. Math. Soc. {\bf
2}(1918)75} \lref\tftads{E. Witten, ``AdS/CFT correspondence and
Topological field theory,'' hep-th/9812012} \lref\jones{E.~
Witten, ``Quantum Field Theory and  the Jones Polynomial", Comm.
Math. Phys. {\bf 121} (1989) 351.} \lref\deboer{J.
de Boer,``Six-dimensional supergravity on $S^3 \times AdS_3$ and 2d
conformal field theory,'' hep-th/9806104; ``Large N Elliptic Genus
and AdS/CFT Correspondence,'' hep-th/9812240}

\lref\carlipbook{S. Carlip, {\it Quantum gravity in 2+1
dimensions}, Cambridge University Press, 1998}
\lref\eguchitaormina{T. Eguchi and A. Taormina, ``Character
formulas for the N=4 superconformal algebra,'' Phys. Lett. {\bf
200B}(1988) 315.}
 \lref\elstrodt{J. Elstrodt, F. Grunewald,
and J. Mennicke, {\it Groups acting on hyperbolic space}, Springer
1998}
 \lref\gottschsoergel{L.
G\"ottsche and W.  Soergel, ``Perverse sheaves and the cohomology
of Hilbert schemes of smooth algebraic surfaces,'' Math. Ann. {\bf
296} (1993)235 }
\lref\hennschwimm{M. Henneaux, L. Maoz, and A. Schwimmer,
``Asymptotic dynamics and asymptotic symmetries of
three-dimensional extended AdS supergravity,'' hep-th/9910013}.

\lref\littlewood{J.E. Littlewood, {\it Littlewood's Miscellany},
Cambridge Univ. Press, 1986} \lref\gritsenko{V. Gritsenko,
``Complex vector bundles and Jacobi forms,'' math.AG/9906191;
``Elliptic genus of Calabi-Yau manifolds and Jacobi and Siegel
modular forms,'' math.AG/9906190}
\lref\borisov{L.A. Borisov and A, Libgober, ``Elliptic Genera
and Applications to Mirror Symmetry,'' math.AG/9904126}

\lref\dmvv{R. Dijkgraaf, G.
Moore, E. Verlinde and H. Verlinde, ``Elliptic Genera of Symmetric
Products and Second Quantized Strings,'' Commun.Math.Phys. 185
(1997) 197-209} \lref\mms{J. Maldacena, G. Moore, and A.
Strominger, ``Counting BPS Blackholes in Toroidal Type II String
Theory,'' hep-th/9903163 }
\lref\zoo{G. Moore and N. Seiberg, ``Taming the conformal zoo,''
Phys. Lett. {\bf 220B} (1989) 422}
 \lref\iwaniec{H. Iwaniec, {\it
Topics in Classical Automorphic Forms}, AMS Graduate Studies in
Math. {\bf 17} 1997} \lref\townsend{A. Achucarro and P.K.
Townsend, ``A Chern-Simons Action for Three-Dimensional
Anti-de-Sitter Theories,'' Phys. Lett. {\bf B180}(1986)89}
 \lref\maldasusskind{Juan M. Maldacena
and Leonard Susskind, "D-branes and fat black
holes",hep-th/9604042, Nucl. Phys {\bf B475}(1996), 679-690}
\lref\petersson{H. Petersson, ``\"Uber die Entwicklungskoeffizienten
der automorphen Formen,'' Acta. Math. {\bf 58}(1932) 169}
\lref\radpaper{H. Rademacher, ``The Fourier coefficients of the
modular invariant $J(\tau)$,'' Amer. J. Math. {\bf 60}(1938)501}

\lref\sarnak{P. Sarnak, {\it Some applications of modular forms},
Cambridge 1990. }

 \lref\witteneg{E. Witten, ``Elliptic Genera and
Quantum Field Theory,'' Commun. Math. Phys. {\bf 109}(1987)525;
``The index of the Dirac operator in loop space,'' Proceedings of
the conference on elliptic curves and modular forms in algebraic
topology, Princeton NJ, 1986.}

\lref\mawa{J. David, Gautam Mandal, S. Vaidya and
Spenta R. Wadia , hep-th/9906112,
Nucl. Phys. {\bf B 564} (2000) 128.}

\lref\ManschotHA{
  J.~Manschot and G.~W.~Moore,
  ``A Modern Farey Tail,''
  arXiv:0712.0573 [hep-th].
}

\Title{\vbox{\baselineskip12pt
\hbox{hep-th/0005003}
\hbox{   }
}}
{\vbox{\centerline{A Black Hole Farey Tail}
 }}
\bigskip
\centerline{Robbert Dijkgraaf, $^1$
Juan Maldacena,  $^{2}$
 Gregory Moore,
$^{3}$ and
Erik Verlinde$^{4}$}
\bigskip
\bigskip
\centerline{${}^1$ Departments of Physics and Mathematics}
\centerline{University of Amsterdam, 1018 TV Amsterdam  }
\bigskip
\centerline{${}^{ 2}$   Department of Physics}
\centerline{Harvard University, Cambridge, MA 02138}
\bigskip
\centerline{${}^3$ Department of Physics and Astronomy, }
\centerline{Rutgers University,
Piscataway, NJ 08855-0849}
\bigskip
\centerline{${}^4$ Joseph Henry Laboratories,}
\centerline{
Princeton University, Princeton NJ 08544  }

\bigskip
\centerline{\bf Abstract}
\vskip .2in
\noindent
We derive an exact expression for the
Fourier coefficients of elliptic genera
on Calabi-Yau manifolds which
is well-suited to studying the AdS/CFT
correspondence on $AdS_3 \times S^3$.
The expression also elucidates an
$SL(2,\IZ)$ invariant phase diagram for
the D1/D5 system involving deconfining
transitions in the $k\rightarrow \infty$
limit.

\Date{}

\newsec{Introduction and Summary}

One of the cornerstones of the AdS/CFT
correspondence
\review\ is the relation between
the partition function $Z_{X}$ of a
superstring theory on $AdS \times X$
and the partition function $Z_\CC$
of a holographically related conformal
field theory $\CC$ on the boundary
$\p (AdS)$. Roughly speaking we
have
\eqn\adscft{
Z_X \sim Z_\CC.
}
While the physical basis for this relationship
is now well-understood, the precise
mathematical formulation and
meaning of this equation has not
been very deeply explored.
This relationship is hard to test since it is difficult
to calculate both sides in the same region of parameter
space.
 In this paper we consider the
calculation of a protected supersymmetric partition function and
we will give an example of
a precise and exact version
of \adscft.
In particular, we will focus on
the example of the duality
between the IIB string
theory on $AdS_3 \times S^3 \times K3$
(arising, say, from the near horizon limit
of $(Q_1,Q_5)$  (D1,D5) branes) and the
dual conformal field theory with
target space ${\rm Hilb}^k(K3)$, the
Hilbert scheme of $k=Q_1 Q_5$ points on
$K3$. We will analyze the so called ``elliptic genus'' which
is a supersymmetry protected quantity and can therefore be
calculated at weak coupling producing a result that is independent
of the coupling.
We will rewrite it in a form which reflects very strongly the
sum over geometries involved in the supergravity side.
We will then give an application of
this formula to the study of phase transitions
in the D1D5 system.

Since the main formulae below are technically
rather heavy we will, in this introduction,
explain the key  mathematical results in a simplified
setting, and then draw an analogy to the physics.
The mathematical results are based on techniques
from analytic number theory, and have their
historical roots in the Hardy-Ramanujan formula
for the partitions of an integer $n$ \hardyram.

Let us consider  a modular form for $\Gamma:=SL(2,\IZ)$
of weight $w<0$  with  a $q$-expansion:
\eqn\introi{
f(\tau)  = \sum_{n\geq 0} F(n) q^{n+ \Delta}
}
where $F(0)\not=0$.
In the physical context
$\Delta = -c/24 $, where $c$ is the central
charge of a conformal field theory,
$w=-d/2 $ if there are
$d$ noncompact bosons in the
conformal field theory, and $F(0)$ is a
ground-state degeneracy.  It is well-known
to string-theorists and   number-theorists alike
that the leading asymptotics of $F(\ell)$ for
large $\ell$ can be obtained \'a la Hardy-Ramanujan
from a saddle-point approximation, and are given by:
\eqn\introii{
F(\ell) \sim {1 \over  \sqrt{2}}  F(0)
\vert \Delta \vert^{(1/4-w/2)  } (\ell+\Delta)^{ w/2 - 3/4}
\exp\biggl[
4\pi  \sqrt{\vert \Delta \vert (\ell + \Delta)} \biggr]
\Biggl(1 + \CO(1/\ell^{1/2}) \Biggr).
}
This estimate
 is a key mathematical
step when accounting
for the entropy of extremal
supersymmetric five-dimensional
black holes  in terms of D-brane microstates
\sv.

What is perhaps less well-known is that there is
an {\it exact} version of the formula \introii,
which
makes the asymptotics manifest.
The Fourier coefficients are given by
the expression:
\eqn\introiv{
\eqalign{
  F(\ell) = 2 \pi \sum_{n+\Delta<0} &
 \biggl({\ell +\Delta \over \vert n+ \Delta \vert}\biggr)^{(w-1)/2} F(n)
\cdot \cr
&
\cdot \sum_{c=1}^{\infty} {1 \over  c} Kl(\ell+\Delta,  n+\Delta;c)
   I_{1-w}\biggl({4 \pi \over  c}
\sqrt{\vert n+ \Delta\vert (\ell+ \Delta)} \biggr).  \cr}
}
The first sum is over the Fourier coefficients
of the ``polar part'' $f^-$ of $f$, defined
by
\eqn\polarpart{
f^-:= \sum_{n+\Delta<0} F(n) q^{n+\Delta}.
}
  In the second sum
  $I_{1-w}$ is the standard
Bessel function and   $Kl(\ell,m;c)$ is the ``Kloosterman sum''
\eqn\kloosterman{
Kl(n,m;c):=\sum_{d \in (\IZ/c\IZ)^*}
\exp\biggl[ 2 \pi i (d {n  \over  c} +  d^{-1}{m  \over  c})\biggr] .
}
The summation variable ``$c$'' in \introiv\ is
traditional. We hope it will not be confused
with the central charge of a conformal field
theory.  Because
$I_\nu(z) \sim z^\nu$ for $z \rightarrow 0$,
the series \introiv\ is absolutely convergent
for $w<1/2$ (see eq. (2.18) below).
On the other hand, in view of the
asymptotics
$I_\nu(z)   \sim \sqrt{1\over  2\pi z} e^{z} $ for
$  Re(z) \rightarrow +\infty$, \introiv\  generalizes
\introii. Finally, to suppress
some complications we will assume in this
introduction (but not in subsequent sections)
that $\Delta$ is a negative integer.
\foot{The formula \introiv\  is due to Rademacher.
The relatively trivial estimate \introii\ is usually
referred to as the Hardy-Ramanujan formula, but in fact,
their full formula is much closer to \introiv. See
especially \hardyram, eqs. 1.71-1.75. See \petersson\radpaper\
for original papers. For some
  history see Littlewood's Miscellany
\littlewood, p. 97,
\apostol, ch. 5, the review article  \knopp, and Selberg's collected works. }

The expression \introiv\ may be usefully
rewritten by introducing the map
\eqn\introiii{
f(\tau) \rightarrow \CZ_f(\tau) := \bigl(q {\p \over  \p q}
\bigr)^{1-w} f
}
We will call this the ``\holographic\ transform.''
 Mathematically it is simply a special case of
Serre duality, but the physical meaning should
be clarified. (We comment on this below.)
In any case, it is in terms of this new modular
form that formulas look
simple.
One readily verifies that, for $w$ integral,  the \holographic\
transform takes a form of modular weight $w$ to a form of   modular
weight $2-w$ with Fourier coefficients  $\tilde F(n) :=
(n+\Delta)^{1-w} F(n)$. Moreover, the transform takes a polar
expression to a polar expression: \eqn\parlpol{ \CZ_f^- = \CZ_{f^-}
= \sum_{n+\Delta<0} \tilde F(n) q^{n+\Delta}. } Notice that $f$ and
$\CZ_f$ contain the same information except for states with
$n+\Delta =0$. Now, using a straightforward application of the
Poisson summation formula (see appendix C), one can cast \introiv\
into the
  form of an average over
modular transformations:
\eqn\introvii{ \CZ_f(\tau) =
\sum_{\Gamma_\infty\backslash \Gamma} (c \tau + d)^{w-2}
\CZ_f^-({a \tau + b \over  c \tau + d} ). }
Here $\Gamma_\infty$ is the subgroup of the modular group $\Gamma$
generated by $\tau \rightarrow \tau+1$. It is necessary to average
over the quotient $\Gamma_\infty\backslash \Gamma$ rather than
$\Gamma$ to get a finite expression, since $\exp[2\pi i (n+\Delta)
\tau]$ is invariant under $\Gamma_\infty$. Note that
$\Gamma_\infty\backslash \Gamma$ can be identified with the set of
relatively prime integers $(c,d)$ or, equivalently, with $d/c \in
\IQ \cup \infty:= \widehat{\IQ}$. \foot{Actually, over two copies of
$\widehat{\IQ}$, if $\Gamma= SL(2,\IZ)$ and not $PSL(2,\IZ)$. This
distinction becomes important if the weight $w$ is odd.}  In
mathematics such averages over the modular group are called
``Poincar\'e series.''

One final mathematical point is needed to complete the circle of
mathematical formulae we will need. Using an integral representation
of the Bessel function one can also write \introiv\ in the form
\eqn\introviii{ \tilde F(\ell) = {1 \over  2 \pi i} \int_{\epsilon -
i \infty}^{\epsilon+ i \infty} e^{2 \pi \beta( \ell + \Delta) }
\widehat \CZ_f(\beta) d \beta } where $\epsilon\rightarrow 0^+$ and
we have introduced the ``truncated sum'' \eqn\introxi{ \widehat
\CZ_f(\beta)   := \sum_{(\Gamma_\infty \backslash
\Gamma/\Gamma_\infty)'} (c \tau + d)^{w-2} \CZ_f^-( {a \tau + b
\over  c \tau + d} ). } Here and throughout this paper $\tau=i
\beta$ in formulae of this type.  When we want to emphasize modular
aspects we use $\tau$, when we want to stress the relation to
statistical mechanics we use $\beta$. Note that $\beta$ is a complex
variable with positive real part. The sum in \introxi\ is a
truncated version of that in \introvii. The prime in the notation
$(\Gamma_\infty \backslash \Gamma/\Gamma_\infty)'$ means we omit the
class of $\gamma=1$. Elements of the double-coset may be identified
with the rational numbers $-d/c$ between $0$ and $1$. For further
details see equations 2.7 - 2.10  below.

We may now describe the physical interpretation
of the formulae \introi\ to
\introxi.  $f(\tau)$ will become
a conformal field theory partition function.
The \holographic transform $\CZ_f(\tau)$
will be the dual supergravity ``partition function.''
The sum
over the modular group in \introvii\
will be a sum over solutions to supergravity.
The \holographic transform is related to
the truncated sum by
\eqn\introvii{
\eqalign{
\CZ_f(\tau) & = \CZ_f^-(\tau)
+ \sum_{\ell\in \IZ} \widehat\CZ_f(\tau +\ell) . \cr}
}
In order to understand this relation,
  recall that
in statistical mechanics a standard maneuver is to
relate the canonical and microcanonical ensemble by
an inverse Laplace transform:
\eqn\microcan{
N(E) = {1 \over   2 \pi i }
\int_{\epsilon - i \infty}^{\epsilon + i \infty}
Z(\beta) e^{2 \pi \beta E} d \beta.
}
Regarding a modular form such as
\introi\ as a partition function
the corresponding microcanonical ensemble
is given by
\eqn\sumens{
N(E) = \sum_{n \in \IZ} \delta(E-n) F(n)
}
and thus we recognize the Rademacher
expansion as the standard
relation between the microcanonical and canonical
ensembles, where the latter is a Poincar\'e series.

Let us pass now from the simplified version
to the true situation.
As we discuss in section two, the formulae \introi\ to
\introxi\ can
be considerably
generalized. In particular,
they  can be applied to Fourier
coefficients of    Jacobi forms of nonpositive
weight, and thus can be applied  to elliptic genera of
Calabi-Yau manifolds. Recall that if $X$ is a
Calabi-Yau manifold then the elliptic genus
may be defined in terms of the associated
$(2,2)$ CFT trace:
 \eqn\ellgen{
\chi( q,y;X) :=
{\Tr}_{RR} e^{2 \pi i \tau  (L_0-c/24) }
 e^{2 \pi i \tilde \tau (\tilde  L_0-c/24) }
 e^{2 \pi i z  J_0}    (-1)^{F }
:=  \sum_{n\geq 0, r } c(n,r) q^n y^r.
}
We will let $q:= e^{2 \pi i \tau}, y:= e^{2\pi iz }$, and
$(-1)^{F }  = \exp[i \pi (J_0 - \tilde J_0)]$.
We collect some standard facts about the
elliptic genus in section three.

As explained in section four, the  analog of the \holographic
transformation  for a Jacobi form of weight $w$ and index $k$ will
be the map \eqn\introv{ \phi \rightarrow FT(\phi):= \biggl\vert
q\p_q - {1 \over  4k} (y \p_y)^2 \biggr\vert^{3/2-w} \phi .}
When $w$ is integral we interpret this as a pseudodifferential
operator: $FT(\phi) = \sum \tilde c(n,\ell) q^n y^\ell$ with
\eqn\introvi{ \tilde c(n,\ell):=\vert n- \ell^2/4k \vert^{3/2-w}
c(n, \ell) . }
Formal manipulations of pseudo-differential operators suggest that
$FT(\phi)$ is a Jacobi form of weight $3-w$ and the same index $k$.
However, these formal manipulations lead to a false result, as
pointed out to us by D. Zagier.  Nevertheless, as we show in section
four, it turns out that for $n-\ell^2/4k>0$ the coefficients $\tilde
c(n,\ell)$ can be obtained as Fourier coefficients of a truncated
Poincar\'e series $\hat \CZ_{\phi}$  defined in equation $(4.6)$
below.

Our main result will be a formula for the \holographic transform of
the elliptic genus $\chi$ for the Calabi-Yau manifold $X= {\rm
Hilb}^k(K3)$. The corresponding truncated Poincar\'e  series
takes the form: \eqn\holotmn{ \CZ_\chi(\beta, \omega) = 2 \pi
\sum_{(\Gamma_\infty \backslash \Gamma)_0 } {1 \over (c \tau + d)^3}
\sum_s \CD(s) \exp\biggl[ - 2 \pi i \Delta_s {a \tau + b \over c
\tau + d} \biggr] \Psi_s({\omega \over c \tau + d} , {a \tau + b
\over  c \tau + d} ) }
The notation is explained in the
following paragraph and a
  more precise version appears in equation
$(5.1)$ below. In section five we interpret this formula physically.
The sum $(\Gamma_\infty \backslash \Gamma)_0$ is the sum over
relatively prime pairs $(c,d)$ with $c>0$ (together with the pair
$(c,d)=(0,1)$.)  We will interpret the average over
$(\Gamma_\infty\backslash \Gamma)_0$ as a sum over an $SL(2,\IZ)$
family of black hole solutions of supergravity on $AdS_3 \times
S^3$, related to the family discussed in \stringexclusion. The sum
over $s$ is a {\it finite} sum over those particle states that do
not cause black holes to form. They can be ``added'' to the black
hole background, and the combined system has gravitational action
$\CD(s)\exp( - 2 \pi i \Delta_s \tau)$ where $\CD(s)$ is a
degeneracy of states. Finally $\Psi_s$ will be identified with a
Chern-Simons wavefunction associated to $AdS_3$ supergravity.

In fact, if we introduce the ``reduced mass''
\eqn\reduced{
L_{0}^{\bot }:= L_{0}-{1\over 4k}J_{0}^{2}-
{k\over 4}
}
then the polar part $\CZ_\chi^-$ of the supergravity partition function
can be considered as a sum over states for which $L_{0}^{\bot }<0$. The
calculations of Cvetic and Larsen \cvl\ show that the area of the
horizon of the black hole and therefore its geometric entropy
is precisely determined by a combination of the
mass and (internal) angular momentum that is identical to $L_{0}^{\bot
}$ (one has $S_{BH}=2\pi\sqrt{ kL_{0}^{\bot }}=\pi
\sqrt{4kn-\ell^{2}}$~ . ) Therefore the truncation of the partition
function to states with $L_{0}^{\bot }<0$ describes a thermal gas of
supersymmetric particles in an AdS background, truncated to those
(ensembles of) particles that do not yet form black holes.
It is with this truncated partition function that contact has been
made through supergravity computations \deboer.

The sum over the quotient $(\Gamma_\infty\backslash \Gamma)_0$ as in
\holotmn\ adds in the black hole solutions. It is very suggestive
that the full partition function can be obtained by taking the
supergravity AdS thermal gas answer and making it modular invariant
by explicitly averaging over the modular group. This sum has an
interpretation as a sum over geometries and its seems to point to an
application of a principle of {\it spacetime} modular invariance.

The relation to Chern-Simons theory makes it particularly clear that in
the AdS/CFT correspondence the supergravity partition function is to be
regarded as a vector in a Hilbert space, rather like a conformal block.
Indeed, the Chern-Simons interpretation of RCFT \jones\ is a precursor
to the AdS/CFT correspondence. This subtlety in the interpretation of
supergravity partition functions has also been noted in a different
context in \tftads. The factor of $(c \tau + d)^{-3}$ in \holotmn\ is
perfect for the interpretation of $Z_\chi$ as a half-density with
respect to the measure $dz\wedge d \tau$.

 To be more
precise, the wave-function is a section of a line-bundle
${\cal L}^{k}\otimes {\cal K}$
where ${\cal K}$ is the holomorphic canonical line bundle over the moduli
space, and $k$ is the level of the CS-supergravity theory. The invariant
norm on sections on the line-bundle ${\cal L}^{k}\otimes {\cal K}$ is given
by
\eqn\normwave{
\exp \Bigl( -4\pi k {({\rm Im} z)^2 \over {\rm Im}\tau }\Bigr)
|\CZ_{\chi }(\tau ,z )d\tau dz |^{2}
}
Assuming that this norm is invariant under modular transformations, one
concludes that $\CZ_{\chi }(\tau ,z)$ is a Jacobi form of weight 3
and index $k.$

One application of \holotmn\ is to the study
of phase transitions as a function of $\tau$ in
the $k \rightarrow \infty$ limit. Since there
are states with $\Delta_s \sim k$ there will
be sharp first order phase transitions as $\tau$
crosses regions in an $SL(2,\IZ)$ invariant
tesselation of the upper half plane.
We explain a proof of this phase structure
in section six.

Finally, we may explain the title of this paper. In
our proof of the result \holotmn\
(see appendix B) the sum over rational numbers
$d/c$ is obtained by successive approximations
by Farey sequences, a technique skillfully
exploited by Rademacher, and going back to
Hardy-Ramanujan.  Only one term in the
sum dominates the entropy of the
$D1D5$ system, the successive terms in
longer Farey sequences  constitute a
tail of the distribution, but this tail
is associated with
a family of black holes. This, then, is our
Black Hole Farey Tail.

\bigskip
{\it Note added for V3, Dec. 8, 2007}: Don Zagier pointed out to us
a serious error in versions 1 and 2 of this paper, namely, in those
versions it was asserted under equation $(3.22)$ that $\CO^{3/2-w}
\phi$ is a Jacobi form of weight $3-w$ and index $k$. This turns out
to be false. In addition to this, it turns out that when one
attempts to convert the truncated   Poincar\'e series $\widehat
\CZ_\phi$ of section 4 to a full Poincar\'e series one gets zero,
and thus the series which one would expect to reproduce $\CO^{3/2}
\chi$ in fact vanishes.  As far as we are aware, the central
formulae $(4.5)$ and $(4.6)$, which involve only the truncated
Poincar\'e series are nevertheless correct. Fortunately, the
physical interpretation we subsequently explain is based on this
truncated series,  so our main conclusions are unchanged.

A recent paper \ManschotHA\  has clarified somewhat the use of the
Fareytail transform, and presented a regularized Poincar\'e series
for $\chi$ rather than $Z_\chi$.

\newsec{The Generalized Rademacher Expansion}

In this section we give a rather general result on the asymptotics
of vector-valued modular forms. It is a slight generalization of
results of Rademacher \rademacherii. See also  \apostol, ch.5, and
\knopp.

Let us suppose  we have a ``vector-valued nearly holomorphic modular
form,''  i.e.,  a collection of functions $f_\mu(\tau)$ which form a
finite-dimensional unitary representation of the modular group
$PSL(2,\IZ)$ of weight $w$. Under the standard generators we have
\eqn\rep{ \eqalign{ f_\mu(\tau+1) & = e^{2\pi i \Delta_\mu}
f_\mu(\tau)\cr f_\mu(-1/\tau) & = (-i \tau)^w S_{\mu\nu} f_\nu(\tau)
\cr} } and in general we define: \eqn\genrep{ \eqalign{
f_\mu(\gamma\cdot \tau)& := (-i (c\tau + d))^w M(\gamma)_{\mu\nu}
f_\nu(\tau) \qquad \gamma
 = \pmatrix{ a & b \cr  c & d \cr}
\cr} }
where, for $c>0$ we choose the principal branch of the logarithm.

We assume the $f_\mu(\tau)$ have
 no singularities for $\tau$ in the upper half plane,
except at the cusps $\IQ \cup i \infty$. We may assume they have an
absolutely convergent Fourier expansion \eqn\collec{ f_\mu(\tau) =
q^{\Delta_\mu} \sum_{m \geq 0} F_\mu(m) q^m \qquad \mu = 1, \dots, r
} with $F_\mu(0)\not=0$ and that the $\Delta_\mu$ are real. We
choose the branch $q^{\Delta_\mu} = e^{2\pi i \tau \Delta_\mu}$.  We
wish to give a formula for the Fourier coefficients $F_\mu(m)$.

We will now state in several forms the
convergent expansion
that gives  the Fourier coefficients
 of the modular forms in terms of data
of the modular representation and the polar
parts $f_\mu^-$.  We assume that  $w\leq 0$.
\foot{The arguments in appendices B,C are only
valid for $w<0$. The extension to $w=0$ was
already known in the 1930's. See \knopp.}

The first way to state the result is
\eqn\radfirst{
F_\nu(n) =   \sum_{m + \Delta_\mu<0} \CK_{n,\nu; m, \mu} F_\mu(m)
}
which holds for all $\nu,n$.
The infinite $\times$ finite matrix $\CK_{n,\nu; m, \mu} $
is an infinite sum over the rational numbers in lowest
terms $0 \leq -d/c < 1$:
\eqn\kernelfn{
\CK_{n,\nu; m, \mu}     = \sum_{0\leq -d/c <  1}
\CK_{n,\nu; m, \mu} (d,c)
}
and for each $c,d$ we have:
\eqn\kernelfnii{
\eqalign{
\CK_{n,\nu; m, \mu} (d,c)  :=
&
 -i  \tilde M(d,c)_{n,\nu; m, \mu}
 \cr
 \int_{1 -i \infty}^{1+ i \infty}
&
d \beta
( \beta c)^{w-2} \exp\biggl[
 2 \pi \bigl( {n + \Delta_\nu \over  c^2}\bigr)
 {1 \over  \beta} - 2 \pi (m + \Delta_\mu) \beta \biggr]\cr}
} The matrix $\tilde M(d,c)_{n,\nu; m, \mu}$ is essentially a
modular transformation matrix and is defined (in equation $(2.10)$
below) as follows.

Let $\Gamma_\infty$ be the subgroup of modular transformations $\tau
\rightarrow \tau + n$. We may identify the rational numbers $0\leq
-d/c<1$ with the nontrivial elements in the double-coset $\gamma\in
\Gamma_\infty\backslash PSL(2,\IZ) /\Gamma_\infty$ so the sum on
$-d/c$ in \kernelfn\ is more fundamentally the sum over nontrivial
elements $[\gamma]$ in $\Gamma_\infty\backslash PSL(2\IZ)
/\Gamma_\infty$. To be explicit, consider a matrix
\eqn\explmtrx{ \pmatrix{a & b \cr c & d \cr} \in SL(2,\IZ) }
where $c,d$ are relatively prime integers. Since
\eqn\mtrxmlt{ \eqalign{ \pmatrix{1 & \ell \cr 0 & 1 \cr} \pmatrix{a
& b \cr c& d \cr} & = \pmatrix{ a+\ell c & b + \ell d \cr c & d \cr}
\cr }.
 }
 the equivalence class in $\Gamma_\infty\backslash \Gamma$ only
 depends on $c,d$. When $c\not=0$ we can take $0\leq -d/c< 1$
 because:
 \eqn\mtrxmlti{
\eqalign{ \pmatrix{a & b \cr c&  d\cr} \pmatrix{1 & \ell \cr 0 & 1
\cr} & = \pmatrix{ a  & b + a \ell   \cr c &   d + c \ell  \cr} \cr}
}
shifts $d$ by multiples of $c$. The term $c=0$ corresponds to the
class of $[\gamma=1]$. It follows that
\eqn\moo{ \tilde M(d,c)_{n,\nu; m, \mu} := e^{ 2 \pi i (n +
\Delta_\nu)(d/c) } M(\gamma)^{-1}_{\nu\mu}e^{ 2 \pi i(m +
\Delta_\mu) (a/c)  } }
 only depends on the class of $[\gamma]\in
\Gamma_\infty\backslash PSL(2,\IZ) /\Gamma_\infty$ because of
\mtrxmlt\mtrxmlti. In such expressions where only the equivalence
class matters we will sometimes write $\gamma_{c,d}$.

Our second formulation is based on the
observation that   the integral in
\kernelfnii\  is
essentially the standard Bessel function
$I_\rho(z)$  with  integral representation:
\eqn\integrl{
I_\rho(z) = ({z \over  2} )^{\rho} {1 \over  2 \pi i }
\int_{\epsilon-i\infty}^{\epsilon+i\infty} t^{-\rho-1}
 e^{(t + z^2/(4t))} dt
}
for $Re(\rho)>0, \epsilon>0$. This function
 has asymptotics:
\eqn\inuasym{
\eqalign{
I_\rho(z) & \sim ({z \over  2} )^{\rho} {1 \over  \Gamma(\rho+1)}
 \qquad  z \rightarrow 0 \cr
I_\rho(z) & \sim \sqrt{1\over  2\pi z} e^{z} \qquad Re(z) \rightarrow +\infty
\cr}
}
Thus, we can define
\eqn\radii{
\tilde I_\rho(z) := ({z \over  2})^{-\rho} I_{\rho}(z)
}
and re-express the formula \kernelfn\ as:
\eqn\raddbcst{
\eqalign{
\CK_{n,\nu; m, \mu}  = + 2\pi
&
\sum_{0 \leq -d/c< 1}    c^{w-2} \tilde M(d,c)_{n,\nu;m,\mu}
 (2\pi \vert m+\Delta_\mu \vert )^{1-w}
  \cr
  \qquad &  \tilde I_{1-w}
 \biggl[ {4\pi\over c} \sqrt{\vert m+\Delta_\mu\vert(n + \Delta_\nu)}
\biggr]\cr}
}

Finally, note that the integral in \kernelfnii\ does not
depend on $d$, so one sometimes separates the
 summation over $d$ defining:
\eqn\radiv{ K\ell(n,\nu,m,\mu;c)  := \sum_{0<-d<c; (d,c)=1} e^{2 \pi
i (n + \Delta_\nu)(d/c) } M(\gamma_{c,d})^{-1}_{\nu\mu}e^{ 2 \pi i(m
+ \Delta_\mu) (a/c)  } } for $c>1$. We will call this
  a generalized Kloosterman sum.
For $c=1$ (the most important case!) we have:
\eqn\radiii{
K\ell(n,\nu,m,\mu;c=1)   = S^{-1}_{\nu\mu}
}
Putting these together we have the third formulation
of  the Rademacher expansion:
\eqn\radi{
\eqalign{
F_\nu(n) = 2\pi
\sum_{c=1}^\infty\sum_{\mu=1}^{r}
&
  c^{w-2} K\ell(n,\nu,m,\mu;c)
\sum_{m+ \Delta_\mu < 0}
F_\mu(m)\cr
(2\pi \vert m+\Delta_\mu \vert )^{1-w}
&
\tilde I_{1-w}
 \biggl[ {4\pi\over c} \sqrt{\vert m+\Delta_\mu\vert(n + \Delta_\nu)}
\biggr] . \cr}
}

The function $\tilde I_\nu(z) \rightarrow 1$
for $z \rightarrow 0$. The Kloosterman sum
is trivially bounded by $c$, so we can immediately
conclude that the sum converges for $w<0$.
(In fact, by a deep result of A. Weil, the
Kloosterman sum for the trivial representation
of the modular group is bounded by $c^{1/2}$.) From the
proof in appendix $B$ it follows that
the series in fact converges to the value of
the Fourier coefficient of the modular form.

We will give two proofs of the above results in appendices B and C.
The first follows closely the method used by Rademacher
\rademacherii\apostol. This proof is useful because it illustrates
the role played by various modular domains in obtaining the
expression and is closely related to the phase transitions discussed
in section six below. The second proof, which  is also rather
elementary, but only applies for $w$ integral,  establishes a
connection with another well-known formula in analytic number
theory, namely Petersson's formula for Fourier coefficients of
Poincar\'e series.

\newsec{Elliptic genera  and Jacobi Forms}

\subsec{Elliptic Genera and superconformal field theory}

The elliptic genus for a $(2,2)$ CFT is defined to be:
\eqn\ellgen{
\chi( q,y) :=
{\Tr}_{RR} e^{2 \pi i \tau  (L_0-c/24) }
 e^{2 \pi i \tilde \tau (\tilde  L_0-c/24) }
 e^{2 \pi i z  J_0}    (-1)^{F }
:=  \sum_{n\geq 0, r } c(n,r) q^n y^r
}
 The Ramond sector
spectrum of $J_0, \tilde J_0$ is integral
for $\hat c$ even, and half-integral  for  $\hat c$ odd
so $(-1)^F = \exp[i \pi (J_0 - \tilde J_0)]$ is
well-defined.
In the path integral we are computing
with worldsheet fermionic boundary conditions:
\eqn\bcs{
\sector{e^{2\pi i z} }{+}  \otimes \sector{+}{+}.
}

We will encounter the elliptic genus for unitary
$(4,4)$ theories. These necessarily have
$\hat c = 2 k$ even integral and  $c=3 \hat c = 6k$.
We choose the $\CN=2$
subalgebra so that $J_0 = 2 J^3_0$
has integral spectrum.

General properties of CFT together
with   representation theory  of $\CN=2$ superconformal theory
show that $\chi(\tau,z )$ satisfies the following
identities.
First,  modular invariance
leads to the transformation laws for $\gamma\in SL(2,\IZ)$:
\eqn\jacformp{
\chi({a \tau + b \over  c \tau + d} , {z \over  c \tau + d})
=  e^{ 2 \pi i k { c z^2 \over  c \tau + d} } \chi(\tau,z)
}
Second, the  phenomenon of  spectral flow is encoded in:
\eqn\jacformpp{
\chi(\tau,z+ \ell \tau + m) = e^{-2 \pi i k (\ell^2 \tau+ 2 \ell z)}
\chi(\tau,z)  \qquad \ell,m \in \IZ
}

\subsec{Jacobi forms}

It was pointed out in \kawai\tkawai\ that
the fundamental identities \jacformp\jacformpp\
define what is known in the mathematical
literature as  a ``weak Jacobi
form of weight zero and index $\hat c/2$.''

\defn\ez . A {\it Jacobi form} $\phi(\tau,z)$ of (integral)
weight $w$ and index $k$ satisfies the identities: \eqn\jacform{
\phi({a \tau + b \over  c \tau + d} , {z \over  c \tau + d}) = (c
\tau+d)^w e^{ 2 \pi i k { c z^2 \over  c \tau + d} } \phi(\tau,z)
\qquad \pmatrix{a& b\cr c & d \cr} \in SL(2,\IZ) } \eqn\jacformi{
\phi(\tau,z+ \ell \tau + m) = e^{-2 \pi i k (\ell^2 \tau+ 2 \ell z)}
\phi(\tau,z) \qquad\qquad\qquad \ell,m\in \IZ } and has a Fourier
expansion: \eqn\jacfour{ \phi(\tau,z) = \sum_{n\in \IZ, \ell\in \IZ}
c(n,\ell) q^n y^\ell } where $c(n,\ell) =0$ unless $ 4nk-\ell^2 \geq
0$.

\defn\ez . A {\it weak Jacobi form} $\phi(\tau,z)$ of weight $w$ and index $k$
satisfies the identities \jacform\jacformi\ and
has a Fourier expansion:
\eqn\jacfour{
\phi(\tau,z) = \sum_{n\in \IZ, \ell\in \IZ} c(n,\ell) q^n y^\ell
}
where $c(n,\ell) =0$ unless $n \geq 0$.

The notion of weak Jacobi form   is defined in \ez, p.104.
In physics we must use weak Jacobi
forms and not Jacobi forms since $L_0-c/24 \geq 0$ in
the Ramond sector of a unitary theory.
In a unitary theory    the $U(1)$
charge $\vert \ell \vert \leq \half \hat c= k$  for topological
states so
$4 nk - \ell^2 = 2 \hat c n -\ell^2 \geq -(\hat c/2)^2$.

%
Thanks to \jacformi\ the coefficients $c(n,\ell)$ satisfy
\eqn\sflwshft{ c(n,\ell) = c(n + \ell s_1 + k s_1^2, \ell+ 2k s_1) }
where $s_1$ is any integer.
Therefore, if  $\ell= \nu + 2k s_0$, with integral $s_0$,
\eqn\shifnt{ c(n,\ell) = c(n- {\ell^2-\nu^2\over 4k}, \nu) = c(n-\nu
s_0 - k s_0^2, \nu)}
Thus we obtain the key point, (\ez, Theorem 2.2), that the expansion
coefficients of the elliptic genus as an expansion in two variables
$q,y$ are in fact given by:
 \eqn\egiii{ c(n,\ell)=
c_\ell(2n\hat{c} - \ell^2) } where $c_\ell(j)$ is extended to all
values $\ell=\mu\ \mod\ \hat{c} $ by $c_\ell(N) = (-1)^{\ell-\mu}
c_\mu(N)$.

It follows that we can give a theta function decomposition to the
function $\phi(\tau,z)$ (\ez\ Theorem 5.1):
\eqn\thetafun{  \phi(\tau,z)  = \sum_{-k+1\leq \nu < k} \sum_{n\in
\IZ} c(n,\nu) q^{n-\nu^2/4k} \theta_{\nu,k}(z,\tau) }
where the sum is over integral $\mu$ for $2k$ even and over
half-integral $\mu$ for $2k$ odd, and where
$\theta_{\mu,k}(z,\tau)$, $\mu=-k+1, \dots, k$
 are theta functions:
\eqn\thetmk{ \eqalign{ \theta_{\mu,k}(z,\tau) & := \sum_{\ell\in
\IZ, \ell = \mu \mod 2k } q^{\ell^2/(4k)} y^{\ell} \cr & =
\sum_{n\in \IZ} q^{k(n+\mu/(2k))^2} y^{ (\mu + 2k n)} \cr} }

In the case of the elliptic genus we have:
 \eqn\egi{ \chi(q,y;Z) =\sum_{\mu=-\hat{c}/2
+1}^{\hat{c}/2} h_\mu(\tau) \theta_{\mu,\hat{c}/2}(z,\tau) }
Physically, the decomposition \egi\
  corresponds to separating out the $U(1)$ current
$J$ and bosonizing it in the standard way $J = i \sqrt{\hat c} \p
\phi$.  Then a basis of chiral conformal fields can be taken so that
\eqn\spltfld{ \CO = \CO_q e^{i q \phi/\sqrt{\hat c}} } where $\CO_q$
is $U(1)$ neutral and has weight $h- q^2/(2 \hat c) $. This weight
can be negative. The remaining ``parafermion'' contributions behave
like: \eqn\egii{ \eqalign{ h_\mu(\tau) & = \sum_{j=-\mu^2 \mod
2\hat{c}} c_\mu(j) q^{j/2\hat{c}} \qquad  1-\hat{c}/2 \leq \mu \leq
\hat{c}/2 \cr & = c_\mu(-\mu^2) q^{-\mu^2/2 \hat c} + \cdots \cr } }
where the higher terms in the expansion have higher powers of $q$.
 Equation  \jacformi\ is physically the statement of
spectral flow invariance.  Recall the spectral flow map \seibschw:
\eqn\spectfl{ \eqalign{
  G_{n \pm a}^\pm   & \rightarrow
G^\pm_{n \pm (a + \theta)} \cr L_0 &
 \rightarrow L_0 + \theta J_0 + \theta^2 {\hat c \over  2}\cr
J_0 & \rightarrow J_0 + \theta \hat c \cr} } leaves invariant the
quantity $2 \hat c L_0 - J_0^2$ for all $\theta$.
%

\subsec{Digression:
Elliptic Genera for arbitrary Calabi-Yau manifolds}

We pause to note a corollary of the Rademacher
expansion  which might prove useful in other
problems besides those discussed in this paper.

One of the primary sources of $(2,2)$ CFT's
are 2d susy sigma models with CY target.
Let $X$ be a CY manifold.  Let $\hat c$ be the complex
dimension. The $\CN=2$ SUSY sigma model on $X$ has
$c=3\hat c$.
The leading coefficient in the $q$-expansion of
$h_\mu$ in \egi\ has a nice
topological meaning \witteneg\kawai:
\eqn\expns{
h_\nu = \chi_{\nu + \hat c/2}(X) q^{-\nu^2/2\hat c} + \cdots
}
where
\eqn\holoeuler{
\chi_p(X) := \sum_{q=0}^{\hat c}  (-1)^{p+q} h^{p,q}(X)
}
is the holomorphic Euler character.

Applying the Rademacher expansion to the
modular forms $h_\mu(\tau)$, we observe that
the  relevant Bessel function $I_{3/2}(z)$ is elementary,
so we have:
\eqn\spleeye{
\eqalign{
\tilde I_{-1/2}(z) & = {2 \over  \sqrt{\pi} z^2} T(z) \cr
T(z) & = e^z (1-1/z) + e^{-z}(1+1/z)\cr}
}
Substituting the above into the
general  Rademacher
series \radi\ we get the formula
for the elliptic genus of an arbitrary
 Calabi-Yau manifold $X$ of complex
dimension $\hat c$:
\foot{An unfortunate clash of notation leads to
three different meanings for ``c'' in this formula!}
\eqn\egrad{
\eqalign{
c(n,\ell ;X) = { \sqrt{\hat c}  \over
2 \hat{c} n -\ell^2 }
&
\sum_{c=1}^{\infty}
\sum_{4km-\mu^2<0}
\sum_{ \mu=-\hat c/2 + 1}^{\hat c/2}  c^{-1/2}
K\ell(n,\nu,m,\mu;c) c(2\hat c m - \mu^2;X)\cr
\vert 2\hat c m - \mu^2
\vert^{1/2}
&
T\Biggl[{ \pi  \over  c}{1  \over   (\hat{c}/2) }
 \sqrt{(\mu^2 - 2 \hat c m)(  2 \hat{c} n -\ell^2)}  \Biggr]\cr}
}
where $\ell=\nu\ \mod\ \hat c$.
(Since we are dealing with elliptic genera
of different manifolds we will use the notation
$c(n,\ell;X)$ when we wish to emphasize the dependence
on the manifold $X$.)

Note that combining with \expns\holoeuler\ one sees that almost
all reference to the variety $X$ has disappeared except for a
finite number of Chern classes. In fact, the elliptic genus
carries no more topological information than the Hodge numbers  as
long as the only terms in the expansion of $h_\nu$ with negative
powers of $q$ are the leading ones. That holds for
\eqn\chatbnd{
2\hat c - \bigl({\hat c \over  2} \bigr)^2 \geq 0 }
or $\hat c \leq 8$.
However, a priori for CY manifolds of $\hat c \geq 9$ the elliptic genus
will generally depend on other topological data besides the Hodge
numbers. \foot{We thank V. Gritsenko for pointing out to us that the
elliptic genus in general depends on more data than just the Hodge
numbers. This statement becomes manifest in view of the Rademacher
expansion. See also \gritsenko.} Using dimension formulas for the space
of Jacobi forms, the above bound has been sharpened in \borisov\ where
it has been shown that Hodge numbers of the Calabi-Yau manifold
determine the elliptic genus only if $\hat c < 12$ or $\hat c=13$. This
paper also contains many explicit computations of the elliptic genus
of Calabi-Yau hypersurfaces in toric varieties.

\subsec{Derivatives of Jacobi forms and \holographic
transforms}

We summarize here some formulae which
are useful in discussing the \holographic
transform of Jacobi forms.

Denote the space of (weak) Jacobi forms of weight $w$ and index $k$
by $J_{w,k}$. Introduce the operator \eqn\ohop{ \CO := {\p \over  \p
\tau} - {1 \over  8 \pi i k} \bigl( {\p \over \p z} \bigr)^2 } One
then easily checks that if $\phi\in J_{w,k}$ then \eqn\ohopii{
\biggl( \CO +  { (w-1/2) \over  2i \Im \tau} \biggr) \phi  }
transforms according the the Jacobi transformation laws of weight
$(w+2)$ and index $k$. By composing operators of this type and
``normal ordering'' one can show that
 \eqn\ohopiii{
\sum_{j=0}^\infty {\Gamma(n+1) \Gamma(n+ w -1/2) \over j!
\Gamma(n+1-j) \Gamma(n+w-\half-j) } \biggl( {1 \over 2 i \Im \tau }
\biggr)^j \CO^{n-j} } maps an element of $J_{w,k}$  to a function
(in general, nonholomorphic) which transforms according to the
Jacobi transformation rules with weight $w+2n$ and index $k$, at
least for $n$ integral.
  On the other hand, for  $w$ integral and $n=3/2-w$ the
expression above simplifies to a single term $\CO^{3/2-w}$, where
the latter should be interpreted as a   pseudodifferential operator.

\defn{(The \holographic transform):} We define the
\holographic transform $FT(\phi)$ of $\phi\in J_{w,k}$ to be
$FT(\phi) :=\vert \CO^{3/2-w}\vert \phi$. We also use the notation
$\widetilde \phi = FT(\phi)$.

Note that if $\phi$ is a weak Jacobi form and we define the polar
part of $\phi$ to be: \eqn\negpart{ \phi^- := \sum_{4kn-\ell^2<0}
c(n,\ell) q^n y^\ell } then $(FT(\phi))^- = FT(\phi^-)$.

As pointed out to us by Don Zagier, it turns out that $FT(\phi)$ is
{\it not} a (weak) Jacobi form. Nevertheless,   it is related to a
truncated Poincar\'e series as we explain in the next section.

\newsec{The Rademacher expansion as a formula in
statistical mechanics}

As we discussed in the introduction, in statistical mechanics
 the canonical and microcanonical ensemble
partition functions are related by
an inverse Laplace transform:
\eqn\microcan{
N(E) = {1 \over   2 \pi i } \int_{\epsilon - i \infty}^{\epsilon + i \infty}
Z(\beta) e^{2 \pi \beta E} d \beta .
}
In this section we cast the Rademacher series into
a form closely related to \microcan.

Consider the first formulation, \radfirst\ to \kernelfnii\
for $(n+\Delta_\nu)>0$.
We can
make the change of variable in \kernelfnii\
\eqn\shiftbeta{
\beta \rightarrow - {n+\Delta_\nu \over  m + \Delta_\mu} \beta
= {n+\Delta_\nu \over \vert m + \Delta_\mu\vert} \beta
}
which is valid as long as it does not shift the
contour through singularities of the integrand.

In the formula below we will find that $Z(\beta)$ has a singularity
at $\beta=0$, so we must have $- {n+\Delta_\nu \over  m +
\Delta_\mu}$ real and positive. Since $(m+\Delta_\mu)<0$  this means
we can only make the change of variable \shiftbeta\ for $(n +
\Delta_\nu) > 0$. The contour deformations are valid in this case
(for $w<2$) so we can write: \eqn\microcanp{ (n+\Delta_\nu)^{1-w}
F_\nu(n)  = {1 \over 2 \pi i } \int_{\epsilon^+ - i
\infty}^{\epsilon^+ + i \infty} \widehat{Z_\nu}(\beta) e^{2 \pi
\beta (n + \Delta_\nu) } d \beta } with \eqn\microcanfinal{
\eqalign{ \widehat{Z_\nu}(\beta)= 2\pi \sum_{0 \leq -d/c< 1 } &
\sum_{m + \Delta_\mu < 0}    \bigl(  c \beta  - i d \bigr)^{w-2}
 \qquad \qquad \cr
M(\gamma_{c,d})^{-1}_{\nu\mu}e^{ 2 \pi i(m + \Delta_\mu) (a/c)  }
\vert m + \Delta_\mu\vert^{1-w} F_\mu(m) & \exp \bigl[ ( 2 \pi)
{\vert m+\Delta_\mu \vert \over c( c \beta -i  d) } \bigr] \cr} }
%
In this equation we can take $c>0$ and since $Re(\beta)>0$ we can
use the principal branch of the logarithm to define $(c \beta - id
)^{w-2}$ when $w$ is non-integral. 

We now use this to  derive
the ``statistical-mechanics''  version of the
Rademacher formula for weak Jacobi forms.
These have an expansion of the form \jacfour\
so we aim to give  a formula of the form:
\eqn\statmech{
\tilde c(n,\ell) = \int_0^1 d\omega e^{-2 \pi i \ell \omega} \int_{\epsilon -i
\infty}^{\epsilon+ i \infty} {d \beta \over  2 \pi i}
 e^{2\pi \beta n}  \widehat \CZ_\phi(\beta, \omega)
} where $\tilde c$ are related to $c$ by the \holographic transform
\introvi. The basic idea of the derivation is to use the
decomposition \egi\ and apply the generalized Rademacher series to
the vector of modular forms given in \egii. After some manipulation
we find: \eqn\jacobicbetterl{ \eqalign{ \widehat
\CZ_\phi(\beta,\omega) = 2 \pi  i (-1)^{w+1} \sum_{0 \leq -d/c< 1 }
&
 \sum_{\mu =-k+1}^k  \sum_{m: 4km-\mu^2 < 0}
\tilde c_\mu(4km-\mu^2)
 \qquad \qquad \cr
\bigl(  c \tau + d  \bigr)^{w-3}
\exp\biggl[4 \pi \vert m -{ \mu^2\over  4k}\vert {1 \over  2i} {a \tau + b
\over  c
\tau + d} \biggr]
 & \exp[- 2\pi i k {c \omega^2 \over  c \tau +d}]
\theta_{\mu,k}({\omega\over  c \tau +d} , {a \tau + b \over  c \tau
+ d}  )\cr} }
Recall that $\tau = i \beta$ in this formula. We will refer to our
result \jacobicbetterl\ as the Jacobi-Rademacher formula.
The most efficient way to proceed here is to work backwards by
evaluating the right-hand-side of \statmech\ and comparing to
\microcanfinal. The reader should note that the meaning of $w$ has
changed in this equation, and it now refers  to the weight of the
Jacobi form: $w(\jacobicbetterl ) = w(\microcanfinal)+\half$. In
particular, for the case of the elliptic genus
$w(\microcanfinal)=-\half$ and $w(\jacobicbetterl )=0$. Finally, we
must stress that the derivation of equation \statmech\ is only valid
for $n-\ell^2/4k>0$.

It is useful to rewrite \jacobicbetterl\ in terms of the slash
operator. In general the slash operator   for Jacobi forms of weight
$w$ and index $k$ is defined to be \ez:
 \eqn\slashop{ \biggl(p
\vert_{w,k} \gamma\biggr)( \tau, z):= (c \tau + d)^{-w} \exp\biggl[
- 2 \pi i k {c z^2 \over  c \tau + d}\biggr] p({a \tau + b \over  c
\tau + d}, {z\over  c \tau +d}   ) }
Using this we can write
\eqn\bozo{ \hat \CZ_\phi =  \sum_{ (\Gamma_\infty \backslash
\Gamma/\Gamma_\infty)_0' } \widetilde \phi^-\vert_{3-w,k} \gamma }
where we define the polar part  to be \eqn\polarpart{ \tilde
\phi^-(\tau, z):= \sum_{4km-\ell^2<0} \tilde c_\ell(4km -\ell^2)
e^{2\pi i m \tau + 2 \pi i \ell z} . }
Here $\Gamma=SL(2,\IZ)$, not $PSL(2,\IZ)$,  the notation
$(\Gamma_\infty \backslash \Gamma/\Gamma_\infty)_0$ means that
$c\geq 0$, and the prime indicates we omit the class of $\gamma=1$.

 As with the full Jacobi form $\phi$ we can use spectral flow to
decompose the polar part   in terms of a finite sum of theta
functions: \eqn\jaciv{ \eqalign{ \tilde \phi^- & :=\sum_{\mu=1}^k
\tilde h_\mu^-  \Theta_{\mu,k}^+\cr \tilde h_\mu^- & =
 \sum_{m: 4km-\mu^2 < 0}
\tilde c_\mu(4km-\mu^2) \exp\biggl[2 \pi i ( m -{ \mu^2\over  4k})
\tau \biggr] \cr \Theta^+_{\mu,k}(z,\tau) & :=
\theta_{\mu,k}(z,\tau) + \theta_{-\mu,k}(z,\tau) \qquad 1 \leq \vert
\mu\vert<k\cr \Theta^+_{k,k}(z,\tau)  & := \theta_{k,k} \qquad \vert
\mu \vert = k \cr} }

As in equations \introvii\ to \sumens\ of the introduction the
relation between the microcanonical and canonical ensemble differs
slightly from the relation between the Fourier coefficients and the
truncated Poincar\'e series  $\widehat \CZ_\phi$. In order to write
the full   canonical partition function we extend the sum in \bozo,
interpreted as a sum over reduced fractions $0\leq c/d < 1$  to a
sum over all relatively prime pairs of integers $(c,d)$ with $c\geq
0$. (For $c=0$ we only have $d=1$.) Let us call the result
$\CZ_{\phi}$. In order to produce a truly modular object we must
also allow for $c<0$, that is, we must sum over
$\Gamma_\infty\backslash SL(2,\IZ)$. Extending the sum in this way
produces {\it zero}, because the summand is odd under $(c,d) \to
(-c,-d)$, and thus we fail to produce a Jacobi form whose Fourier
coefficients match those of $FT(\phi)$, for $n- \ell^2/4k
>0$. This is just as well, since, as pointed out to us by Don
Zagier, $FT(\phi)$ is not a Jacobi form.


\newsec{Physical interpretation in terms of IIB
string theory on $AdS_3 \times S^3 \times K3$ }

Let us apply the the Jacobi-Rademacher formula
to the elliptic genus  $\chi$
for ${\rm Sym}^k(K3)$. In this case
$w=0$ and \bozo\ becomes
\eqn\morphys{
\eqalign{
  \CZ_\chi(\beta,\omega) =
-2 \pi i \sum_{(c,d)=1, c\geq 0 }  &
 \sum_{\mu = 1}^k  \sum_{ 4km-\mu^2 < 0}
\tilde c_\mu\bigl( 4km-\mu^2; {\rm Sym}^k(K3)\bigr)
 \qquad \qquad \cr
\bigl(  c \tau + d  \bigr)^{-3}
\exp\biggl[2 \pi i \bigl( m -{ \mu^2\over  4k}\bigr)
 {a \tau + b
\over  c
\tau + d} \biggr]
 & \exp[- 2\pi i k {c \omega^2 \over  c \tau +d}]
\Theta^+_{\mu,k}({\omega\over  c \tau +d} ,
{a \tau + b \over  c \tau + d}  )\cr}
}

We claim that \morphys\   has the interpretation as a
sum over Euclidean geometries in the effective
supergravity obtained by reducing IIB string
on $AdS_3 \times S^3 \times K3$.
The reasoning is the following:
We begin with the   trace for the
$(4,4)$ CFT with target space ${\rm Sym}^k(K3)$:
\eqn\cfttra{
Z_{RR} (\tau,  \tilde \tau, \omega, \tilde \omega) =
{\Tr}_{RR} e^{2 \pi i \tau  (L_0-c/24) }
 e^{2 \pi i \tilde \tau (\tilde  L_0-c/24) }
 e^{2 \pi i \omega  J_0}  e^{- 2 \pi i \tilde \omega \tilde J_0}
(-1)^F }
 As is standard, this trace can be represented as a
partition function of the $(4,4)$ CFT on the torus. $\tau$
specifies the conformal structure of the torus relative to a
choice of homology basis for space and time, while the  fermion
boundary conditions relative to this basis are \eqn\bcs{
\sector{e^{2\pi i \omega } }{+} \otimes \sector{e^{2\pi i \tilde
\omega  } }{+} } By spectral flow $\omega \rightarrow \omega +
\tau/2$ we may relate this to $Z_{NSNS }$ with boundary conditions
\eqn\bcsp{ \sector{e^{2\pi i \omega } }{-} \otimes \sector{e^{2\pi
i \tilde \omega  } }{-} }
the precise relation being
\eqn\mapsectors{ Z_{RR}(\tau, \omega; \tilde \tau, \tilde \omega)
=
(q \bar q)^{k/4} y^k\tilde y^k Z_{NSNS}(\tau, \omega + \half \tau;
\tilde \tau, \tilde \omega + \half \tilde \tau) }
where $Z_{NSNS}$ is defined as in \cfttra\ in the NSNS sector\foot{
Note that a complex $\omega$ in \bcs\ is equivalent to inserting a
phase $e^{2\pi i \omega_1}$ in the vertical direction and a phase
$e^{2\pi i \omega_2}$
in the horizontal direction with $\omega = \omega_1 + \tau \omega_2$,
where $\omega_1, ~ \omega_2$ are real. In other words
$\sector{e^{2\pi i \omega  } }{+} =
\sector{e^{2\pi i \omega_1 } }{ e^{ 2 \pi i \omega_2 } }$}.
Note that $\omega$ is inserted relative to $(-1)^F$.

Now we use the AdS/CFT correspondence \review. The conformal field
theory partition function
 is identified with a IIB
superstring partition function. The reduction on $AdS_3 \times S^3
\times K3$ leads to an infinite tower of massive propagating
particles and a ``topological multiplet'' of extended $AdS_3$
supergravity \degersezgin\deboer. The latter is described by a
super-Chern-Simons theory \townsend, in the present case based on
the supergroup $SU(2\vert 1,1)_L \times SU(2 \vert 1,1)_R $
  \deboer.
The modular parameter $\tau$ and the twists in \bcs\ are
specified in the supergravity partition function through the
boundary
conditions on the fields in the super Chern-Simons theory. These
involve the metric, the $SU(2)_L \times SU(2)_R$ gauge fields, and
the gravitini. The boundary conditions are as follows:

\item{1.}
The path integral over 3d metrics will involve a sum over
asymptotically  hyperbolic geometries which bound the torus of
conformal structure $\tau\ \mod\ SL(2,\IZ)$.  Thus, we  sum over
Euclidean 3-metrics with a conformal boundary at $r=\infty$ and
\eqn\metricasymp{ ds^2 \sim {dr^2 \over r^2} + r^2 g_{ij} dx^i
dx^j,}
 where $g_{ij}$ is in the conformal class of a torus with
modular parameter $\tau$, and $r$ is a radial coordinate near the
conformal boundary.

\item{2.} Using the known relation between $SU(2)$
 Chern-Simons theory and boundary current algebra
 \jones, in particular, using the evaluation of the
 wavefunction on the torus given in \emss, we see that the
 CFT trace with and insertion of $\exp[2\pi i \omega J_0]$
 entails boundary conditions on $A\in su(2)_L$:
 \eqn\aleftgf{
 A_u du  \rightarrow {\pi \over 2 \Im \tau} \omega \sigma^3 du
 }
 Note that in Chern-Simons theory we specify boundary
 conditions on $A_u$, but leave $A_{\bar u}$ undetermined.
Similarly, we have:
\eqn\arightgf{
 \tilde A_{\bar u} d\bar u
 \rightarrow {\pi \over 2 \Im \tau}\tilde  \omega \sigma^3
 d\bar u
 }

\item{3.} Finally we must choose boundary conditions for the
spinors, in particular for the gravitini in the fermionic part of
the $SU(2\vert 1,1)_L \times SU(2\vert 1,1)_R$ connection. (These
become the supercurrents in the boundary CFT.)  Since these
fermions are coupled to the $SU(2)_L \times SU(2)_R$ gauge fields
the fermion conditions can be shifted by turning on a flat
connection. If the boundary conditions on the gauge fields are
given by \aleftgf\arightgf\ and  we wish to compute $Z_{RR}$, then
the fermion boundary conditions should be
as in \bcs , but since we wish to  specialize to
the elliptic genus then  we must
put $\tilde \omega=0 $ in \bcs, leaving us with
\eqn\bcspp{ \sector{e^{ 2 \pi i \omega}  }{+} \otimes \sector{ + }{+} .}
%

There are many geometries that contribute to this partition function.
Let us first start by discussing the simplest ones. The simplest of these
geometries   are the ones which can be obtained as solutions of the
$SU(2|1,1)^2$ Chern Simons theory. This Chern Simons theory is a consistent
truncation of the six dimensional supergravity theory. So a solution
of the Chern Simons theory will also be a solution of the six dimensional
theory. These solutions correspond to choosing a way to fill in
the torus. This corresponds to picking a primitive one cycle $\gamma_r$
and filling in the torus so that $\gamma_r$ is contractible.
The $U(1)_L$ gauge connection is  flat; in a suitable gauge it
 just a given by  constant $A_u$ and $A_{\bar u}$. As we said above
the constant value of $A_u$ corresponds to the parameter $\omega$
in the partition function through \aleftgf . In the classical
solution $A_{\bar u}$ is determined by demanding that the full
configuration is non-singular. This translates into the condition
that the Wilson line for a unit charge particle around the
contractible cycle is minus one, in other words
\eqn\wline{
  e^{i \int_{\gamma_r} A } =-1 .}
This ensures that we will have a non-singular solution because
the particles that carry odd charge are fermions which in the absence
of a Wilson line were periodic around $\gamma_r$. With this particular
value of the Wilson line they are anti-periodic, but this is precisely
what we need since the geometry near the region where $\gamma_r$ shrinks
to zero size looks like the origin of the plane.
All that we have said for the $U(1)_L$ gauge field should be repeated
for the $U(1)_R$ gauge field. Since there are fermions that carry
charges $(1,0)$ or $(0,1)$ we get the condition \wline\ for both
$U(1)_{L,R}$ connections.
In this way we resolve the paradox that the $(++)$ spin structure
in \bcspp\ cannot be filled in.
For more details on these solutions in the Lorentzian
context, see
\ref\mmaoz{
J.~M.~Maldacena and L.~Maoz,
``De-singularization by rotation,''
JHEP {\bf 0212}, 055 (2002)
[arXiv:hep-th/0012025];
V.~Balasubramanian, J.~de Boer, E.~Keski-Vakkuri and S.~F.~Ross,
``Supersymmetric conical defects: Towards a string theoretic description  of black hole formation,''
Phys.\ Rev.\ D {\bf 64}, 064011 (2001)
[arXiv:hep-th/0011217].
}.
Note that there is an infinite family of solutions that solves
\wline\ since we can always add a suitable integer to $A_{\bar u}$.
This corresponds to doing integral units of spectral flow.
For the purposes of this discussion we can take
the $k\to \infty$ limit, and in this limit  one can examine
the  {\it classical} equations, and hence
 specify the values of both $A_{u}$ and $A_{\bar u}$.
Note that the final effective
boundary conditions of the supergravity fields
depends   on both $A_u $ and $A_{\bar u}$. In particular, they
are not purely given in terms of the field theory boundary conditions
(which is the information  contained in  $A_{u}$).\foot{
Note that this geometry is basically Euclidean $AdS_3 \times S^3$ with
an identifiction in the time direction. This is sometimes loosely
refered to as the ``NS vacuum''. Nevertheless it also corresponds
to a particular RR vacuum, the one with maximal angular momentum
\mmaoz . {} From the boundary field theory point of view we know that
the NS sector and the RR sector are related by a simple spectral flow
transformation that only changes the $U(1)$ charge of the state. The
bosonic field corresponding to the bosonized $N=2$ U(1) currrent on the
boundary is  a singleton living on the boundary of $AdS$. So
configurations in the boundary theory which only differ by the charge
under this $U(1)$ are identical in the interior of $AdS$.}
The final boundary
conditions for the fermions in the supergravity theory depend also on the
particular state that we are considering.
The simple solutions that we have been discussing
correspond to the   $m=0$ and $\mu=k$ term
in \morphys . The sum over all possible contractible cycles corresponds
to the sum over $c,d$ in \morphys . And the sum over integer
values of spectral flow corresponds to the different terms in
the sum over integers in the theta function in \morphys .
Note that from the point of view of an observer in the interior all
these solutions are equivalent, (up to a coordinate transformation),
to Euclidean $(AdS_3 \times S^3)/\IZ$. There is, however, nontrivial
information in this sum since we saw that it is crucial for recovering
the full partition function of the theory.

Now that we have discussed the simpest solutions we can ask about all
the other terms, i.e. about the sum over $m, \mu$ in \morphys .
These correspond to adding particles to the solutions described in
the above paragraph.  These particles are not contained in the
Chern Simons theory. The six dimensional theory, reduced on
$S^3$ gives a tower of KK fields that propagate on $AdS_3$.
If we compute the elliptic genus for them only a very small subset
contributes. From the point of view of the Chern Simons theory adding
these particles is like adding Wilson lines for the $U(1)$
connection  \zoo\emss. \foot{ Note that the Chern Simons description only makes
sense for distances much larger than the $AdS$ radius since some
of the particles in question have compton wavelengths of the order
of the AdS radius. }
In this case we do not have the relation  \wline\ near the boundary.
This is not a problem since the connection is not flat any longer
in the full spacetime. It is possible to find complete non-singular
six (or ten)  dimensional solutions which correspond to
 various combinations of $RR$ ground states \ref\LuninIZ{
O.~Lunin, J.~Maldacena and L.~Maoz,
``Gravity solutions for the D1-D5 system with angular momentum,''
arXiv:hep-th/0212210.
}.

In the next several sections we discuss the physical interpretation of
various aspects of the formula \morphys\ to justify the claim that
\morphys\ takes exactly the form expected from an evaluation of the
partition function of type IIB string theory on $AdS_3 \times S^3 \times
K3$.

\subsec{Interpreting the sum over $d/c$}

We interpret the sum over relatively prime integers $(c,d)$ as a sum
over the ``$SL(2,\IZ)$ family of black holes'' discussed by Maldacena
and Strominger in \stringexclusion. Since this subject is apt to cause
confusion we will be somewhat pedantic in this section,
where we explain why we sum over $(c,d)$ and not all of $SL(2,\IZ)$.

Recall that we can
  identify Euclidean $AdS_3$, denoted by
$\IH$,  as the space
of Hermitian matrices
\eqn\sott{
{\bf X} = {\ell}^{-1} \pmatrix{ T_1 + X_1 & X_2 + i T_2 \cr
X_2 - i T_2 & T_1 - X_1 \cr}
}
with $X_i,T_i$ real and $\det {\bf X}=1$. Here $\ell$ is the $AdS$ radius,
for simplicity we usually choose units where $\ell =1$.

We introduce global variables on $\IH$
via  the Gauss decomposition
\eqn\gauss{
 {\bf X} =
 \pmatrix{ h + w \bar w/h & w/h \cr
\bar w/h & 1/h \cr} = \pmatrix{1 & w \cr 0 & 1 \cr}
\pmatrix{ h & 0 \cr 0 & h^{-1} \cr} \pmatrix{1 & 0 \cr \bar w & 1\cr}
}
Here $h\not=0$, and $  \bar w \in \IC$ the
complex conjugate of $w$.  Since $(T_1 \pm  X_1) \not=0$ we can always
solve for $h, w, \bar w$ so these
coordinates cover $\IH$ once. There are two
connected components of $\IH$
  and we restrict attention to  the
connected component defined by $h>0$.
The metric becomes:
\eqn\sotiii{
ds^2 = { 1 \over  h^2} (dw  d \bar w  + dh^2)
}
which is the standard model of hyperbolic space.

We now study the BTZ group action
\btz. Abstractly, this is just
  an action of the additive group $\IZ$ on
$\IH$. A generator acts as
\eqn\adsgas{
{\bf X} \sim \pmatrix{ e^{-i \pi \tau} & 0 \cr
0 & e^{i \pi \tau} \cr}
{\bf X}
\pmatrix{ e^{  i \pi \bar \tau} & 0 \cr
0 & e^{- i \pi \bar \tau} \cr}
:=\rho(\tau)  {\bf X}\rho(\tau)^\dagger
}
where we must choose $\Im \tau>0$ for
a properly discontinuous action.
The BTZ group acts as isometries
 in  the hyperbolic metric. In the
coordinates \gauss\
 the BTZ group action \adsgas\  is
\eqn\poiniden{
\eqalign{
w   \sim q  w \qquad & \qquad
h    \sim e^{- 2\pi (\Im \tau)} h \cr}
}
From
\poiniden\  it is now clear that the group action
is not properly discontinuous at $w=0,h=0$. Therefore,
we restrict to the domain  $\IH^*$ defined by
$h>0$ and $h=0, w\not=0$ and take the
quotient by $\IZ$.
The quotient space $\IH^*/\IZ $
is a solid torus with a boundary two-torus:
The identification $w \sim q w$ in $\IC^*$
defines a torus, then  $h>0$   fills it in.

We claim the the modular parameter of
the torus at the conformal boundary $h=0$
is naturally given  by $\tau$, up to an ambiguity
$\tau \sim \tau + \IZ$.
 Recall that the modular parameter of a torus
is only defined  up to a  $PSL(2,\IZ)$ transformation
until one chooses an {\it oriented}
homology basis. Once we choose $a$
and $b$-cycles (in that order!) we can define
$\tau := \int_b \omega / \int_a \omega$
where $\omega$ is a globally nonvanishing
holomorphic 1-form. When we are presented
with a solid torus then there is a unique primitive
contractible cycle. We take this to be the $a$-cycle.
There is no unique noncontractible primitive
cycle; any
two differ by an integral multiple of the $a$-cycle.
 These
different choices are related by diffeomorphisms
of the solid torus which become Dehn twists about
the $a$-cycle on  the boundary. Thus, a choice of
filling in the torus to a solid torus defines a
modular parameter up to $\tau \sim \tau + \IZ$.
Now consider the quotient geometry $\IH^*/\IZ$.
{}From \poiniden\ the unique primitive  contractible cycle is
$w(s) = w_0 e^{2 \pi i s}$, $h(s) = h_0$.
(Note that by making $h_0$ large we can make the length
arbitrarily small.) One choice of noncontractible cycle is
$w(s) = \exp[- 2 \pi i \tau s] w_0$,
joining $w_0$ to $q^{-1} w_0$.  With this
choice of b-cycle the modular parameter is $\tau$.

Now we would like to make a connection to the
physics of black holes and thermal AdS. To
begin we use the   decomposition
\eqn\eusotti{
\eqalign{
{\bf X} & = \pmatrix{e^{u } & 0 \cr 0 & e^{-u}\cr}
\pmatrix{\cosh \rho &\sinh \rho \cr
\sinh \rho & \cosh \rho \cr}
\pmatrix{e^{\bar u} & 0 \cr 0 & e^{-\bar u}\cr}  \cr
}
}
where $\rho \geq 0$ and $u\in \IC$.
Comparing to \gauss\ we have
$w= e^{2 u } \tanh \rho  $.
In these coordinates the metric takes the
form
\eqn\eucpropco{
ds^2 = - \sinh^2\rho  (du - d\bar u )^2 +
\cosh^2\rho
(du + d \bar u  )^2
+  d\rho^2
}
These coordinates cover $\IH^*$ once, and
therefore give global coordinates if we identify
\eqn\concyc{
2u  \sim 2u + 2\pi i n \qquad n\in \IZ
}
By the above remarks, $u(s) = i \pi s$, $0\leq s\leq 1$ is the
unique contractible primitive cycle.
On the other hand, the
BTZ group action identifies
\eqn\btzcyc{
2u \sim 2u + 2 \pi i n \tau \qquad n\in \IZ
}
and defines one (of many) primitive
noncontractible cycles. We will refer to
\btzcyc\ as the ``BTZ cycle.''
The metric at $\rho\rightarrow \infty$ is $ds^2 \sim e^{2\rho} \vert du \vert^2
+ (d \rho)^2$ and, with respect to the homology
basis (contractible cycle, BTZ cycle) the modular
parameter is $\tau$.

So far we have only done geometry and no physics.
The physics comes in when we introduce coordinates
we want to call ``space'' and ``time.'' There is a {\it unique}
complete, smooth, infinite volume hyperbolic
3-manifold with conformal boundary a single
torus with Teichm\"uller parameter
$\tau$ \birmingham. Nevertheless, for physics
we must identify space and time, and physical
quantities such as the gravitational action discussed
in section 5.3 below  are {\it not} invariant
under global diffeomorphisms.

Suppose we define
\eqn\periyou{
2u = i(\phi + i t_E)
}
with
$\phi, t_E$ real. The notation suggests that $\phi$ is a spatial
angular coordinate and $t_E$ is a Euclidean time.
The identification \concyc\ becomes
\eqn\adsgsid{
\eqalign{
\phi & \sim \phi + 2\pi n,\cr
t_E & \sim t_E, \qquad n \in \IZ \cr}
}
The BTZ group action \btzcyc\ becomes the
identification of Euclidean time with a spatial
twist:
\eqn\adsgdii{
\eqalign{
\phi & \sim \phi + 2\pi n \tau_1,  \cr
t_E & \sim t_E + 2 \pi n \tau_2,  \qquad n \in \IZ \cr}
}
The spatial ($\phi$) cycle \adsgsid\  is
the unique primitive contractible cycle.
Substituting   \periyou\ into
\eucpropco\ we recognize the Euclidean
thermal AdS.

On the other hand, we could instead
define coordinates on $\IH^*/\IZ$ using \eusotti:
\eqn\firstper{
2u:=- (\tau_2 \phi + \tau_1 t_E) + i (  \tau_1 \phi - \tau_2 t_E) =
+i \tau (\phi+ i t_E)
}
 In these coordinates the
 identification \concyc\ is equivalent to
\eqn\firstidenp{
\eqalign{
\phi & \sim  \phi + 2 \pi \Re(-1/\tau) n \cr
t_E & \sim  t_E + 2 \pi \Im(-1/\tau) n \qquad n \in \IZ
\cr}
}
This defines  the unique primitive
{\it contractible} cycle in the solid
torus. We call it the time cycle.
The BTZ action \btzcyc\ becomes
\eqn\seciden{
\eqalign{
\phi & \sim \phi + 2 \pi n,  \cr
t_E & \sim t_E, \qquad n \in \IZ \cr}
}
This defines a choice of noncontractible cycle within the
handlebody. We call it the ``space cycle.''
Note that it is the {\it spatial } cycle
\seciden\  which is a noncontractible
cycle: Thus we have a  black hole
since we have a hole in space.
Indeed, identifying $t_E$ as Euclidean time
and $\phi$ as an angular coordinate we
can define the Schwarzschild coordinate $r$ via
\eqn\rhoarr{
\sinh^2 \rho = {r^2 - \tau_2^2 \over  \vert \tau \vert^2}
}
with $r \geq \tau_2$ to get the familiar
Euclidean BTZ black
hole in Schwarzschild coordinates:
\eqn\eucbtz{
\eqalign{
ds^2 & = N^2(r) dt_E^2 + N^{-2}(r) dr^2 + r^2 (d \phi + N^\phi(r) dt_E)^2 \cr
N^2(r)  & = {(r^2 - \tau_2^2)(r^2 + \tau_1^2) \over  r^2} \cr
N^\phi & = + {\tau_1 \tau_2 \over  r^2} . \cr}
}

Note that, in the second description, if we choose
as oriented homology basis (space cycle, time cycle)
then the modular parameter is $-1/\tau$. In this
way thermal AdS  is related to the BTZ black hole
by a modular transformation.

The general story is the following (we are switching
here from a passive to an active viewpoint): We wish to find a
complete hyperbolic 3-geometry with

\item{1.}  $ds^2 \rightarrow r^2 \vert d \phi + i d t \vert^2 + {dr^2 \over
 r^2} $ at $r \rightarrow \infty$.

\item{2.} Periodicities $(\phi + i t) \sim
(\phi + i t)  + 2 \pi (n + m \tau)$, $n,m\in \IZ$.

\item{3.} The unique primitive contractible cycle is
defined by $\Delta(\phi + i t) = c \tau + d$,
where $(c,d)=1$.
\foot{One must include the special cases $(c=0,d=1)$, $(c=1,d=0)$. }

The solution is to take the BTZ group action with $\rho({a \tau +
b \over  c \tau + d} )$ in \adsgas, where
$$\pmatrix{a& b\cr c & d\cr}\in SL(2,\IZ), $$

 and
to define coordinates
\eqn\gencasei{
2u = {i \over  c \tau + d} (\phi + i t)
}
Then the contractible cycle is
$\Delta( \phi + i t) = 2 \pi (c \tau + d) $ and the
BTZ cycle is
$\Delta( \phi + i t) = 2 \pi (a \tau + b) $
so, with respect to the homology basis
(contractible cycle, BTZ cycle) the
modular parameter is ${a \tau + b \over  c \tau + d}$.
The metric is \eucpropco. Substituting
\gencasei\ into
\eucpropco\ we may bring the metric to the BTZ
form \eucbtz\ with $\tau \rightarrow \tau''$ where
\eqn\gencasii{
\eqalign{
r^2  & = {(c \tau_1 + d)^2 \sinh^2 \rho + (c \tau_2)^2 \cosh^2 \rho \over
\vert c \tau + d \vert^4} \cr
\tau''  & = \pm {1 \over  c \tau + d} \cr}
}
where the sign in the second line
 is determined by $\Im \tau'' >0$.

 Finally,  we may attempt to interpret the constraint $c \geq 0$ as
 follows: Modular transformations with $(c,d)$ and $(-c,-d)$ differ
 by the transformation $\gamma=-1$. While this is an orientation
 preserving transformation of the boundary, it can only be extended
 as a diffeomorphism of the handlebody if it is extended as an
 orientation reversing diffeomorphism. For some reason, which should
 be more clearly explained, we only sum over bounding geometries
 with fixed induced orientation.

The above construction of the ``$SL(2,\IZ)$ family of black holes''
of \stringexclusion\ shows that the family is  perhaps more
accurately described as a $(\Gamma_\infty\backslash \Gamma)_0$
family of choices of contractible cycle for the torus at infinity.
(The subscript $0$ indicates we only keep $c\geq 0$.)  The
geometries are really labelled by a pair of relatively prime
integers $(c,d)$ telling which cycle of the torus at infinity should
be considered to be the contractible cycle.

%
%

\subsec{Interpreting the sum on $\mu, m$}

 We now come to the physical
interpretation of the sum over quantum numbers $(m,\mu)$ with
$4km-\mu^2<0$.  We will interpret the sum on $\mu$ as arising from
black holes which are spinning in the internal $S^3$ directions.

\subsubsec{CFT description}

 First, let us clarify the meaning of
the sum on $(m,\mu)$ from the CFT viewpoint. It is useful to look
at these quantum numbers in both the NS and R sectors. The two
descriptions are related, of course, by spectral flow.

\ifig\rstates{The shaded region contains the
points $(m,\ell)$ contributing to the Jacobi-Rademacher
formula.   There are several points in this
region.}{\epsfxsize3.0in\epsfbox{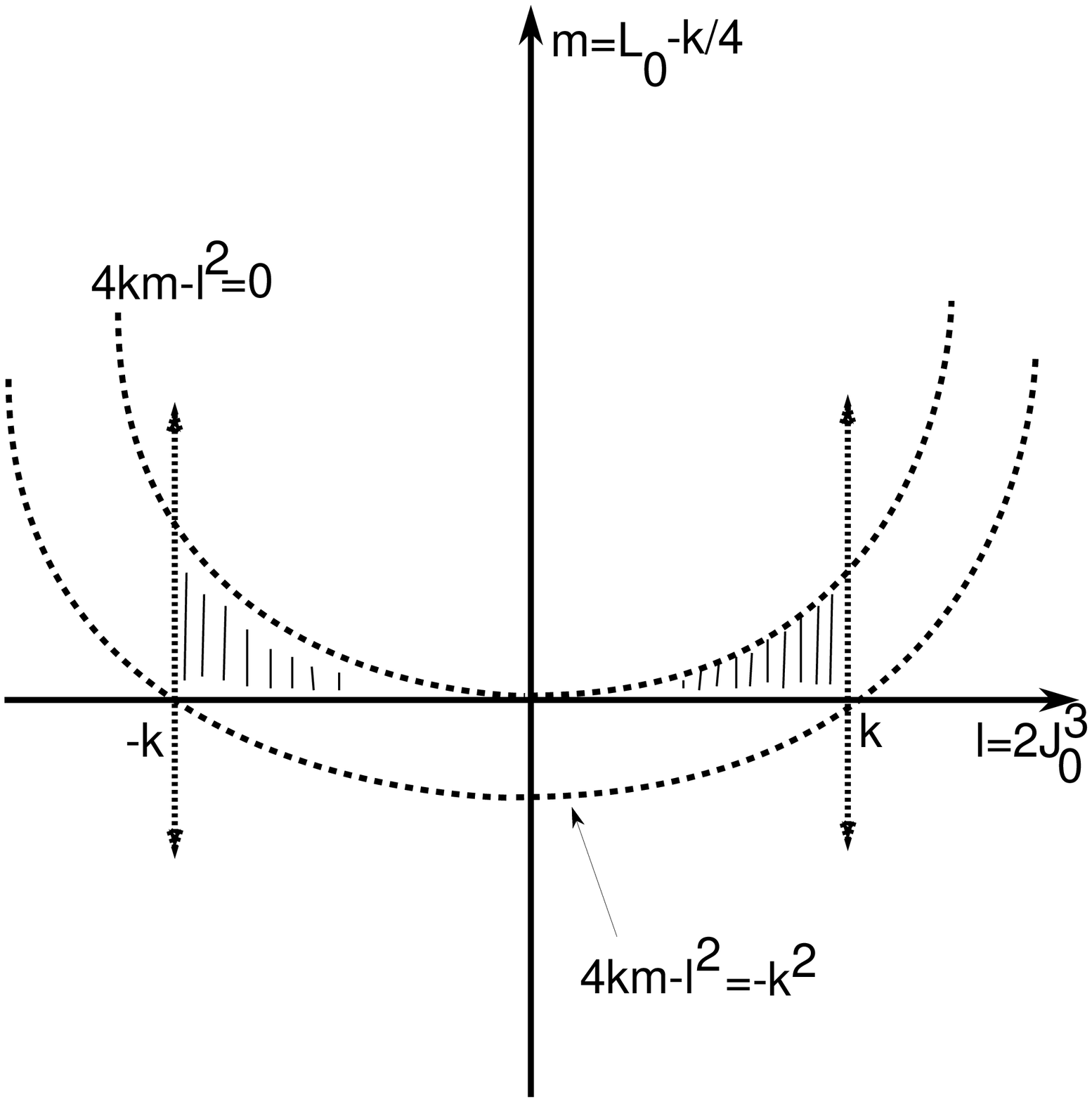}}

\ifig\nsstates{The shaded region is the NS sector particle region.
It is obtained from   \rstates\  by spectral
 flow by $\theta=+\half$. The straight line describes
the left chiral primary states, and all other states of the theory
lie above this line. The parabola describes the cosmic censorship
bound for black holes rotating in the $S^3$ directions. All black
holes lie on or above the parabola. There is an extremal black
hole state that is also a chiral primary with angular momentum
$J_0 = k $. }{\epsfxsize3.0in\epsfbox{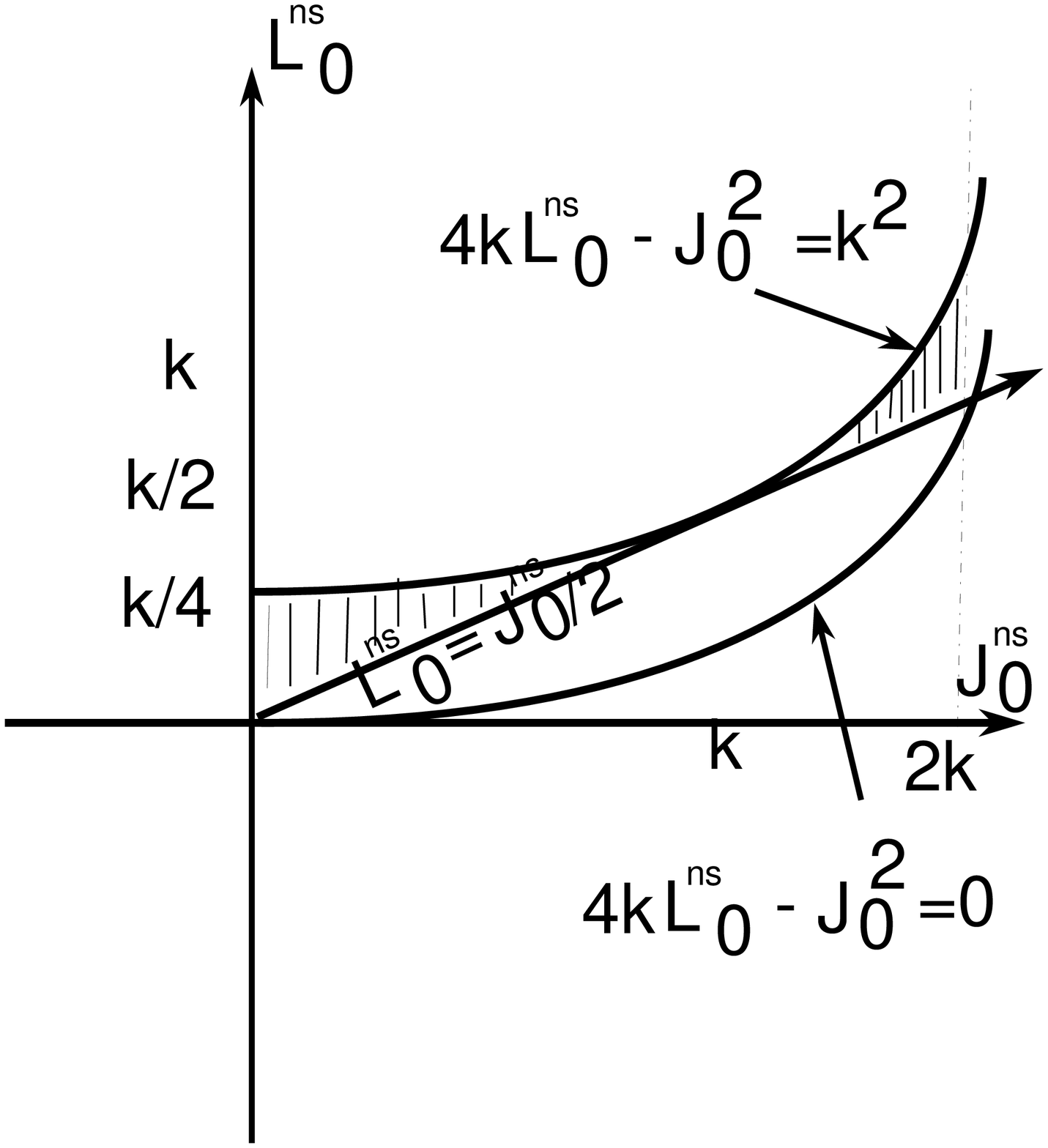}}

In the Ramond sector we plot $m= L_0 - k/4$ against $\ell$ in
\rstates. The key region is: $m\geq 0, \vert \ell\vert \leq k$,
$-k^2 \leq 4km-\ell^2 < 0$, and we will refer to this as the
``Ramond sector particle region,'' for reasons which will become
clear in a moment. Physical states correspond to integral values
of $m,\ell$.  All the topological states with $m=0$ contribute,
{\it except} for $(m=0,\ell=0)$. From the $\CN=4$ character formulae
of \eguchitaormina\ one can check   that other states besides just
the $\CN=4$ descendents of topological states at $m=0$ will
contribute to the elliptic genus.

Now let us compare things in the NS sector. Recall $4kL_0 - J_0^2$
is a spectral flow invariant. One should be careful to distinguish
$m$ from $L_0$ since they differ by $m=L_0 - k/4$ so that
\eqn\spfbds{ 4km - \ell^2 = 4 k L_0 - J_0^2 - k^2.  }
The region in the NS sector is obtained by spectral flow and is
illustrated in \nsstates. For flow by $\theta=+\half$ we get:
\eqn\nsregion{ \eqalign{ 0 \leq 4k L_0^{NS} - (J_0^{NS})^2 < k^2
\cr 0\leq  J_0^{NS}   \leq 2 k \cr} }
Physical states satisfy $(J_0,L_0 )\in \IZ\times \half \IZ_+$ with
$L_0-\half J_0\in \IZ_+$.

It is interesting to see which chiral primary points $(J_0,L_0)$
contribute to the sum in the Jacobi-Rademacher series.  For a
chiral primary $L_0 = J_0/2$ so the spectral flow invariant
\spfbds\ becomes $- (J_0-k)^2$. Thus, all but the middle $\CN=2$
chiral primaries  with $J_0 = k$ contribute.  The point $(J_0,L_0)=
(k, \half k)$ is a distinguished point. These are the quantum
numbers of a state which  is both a chiral primary and - as we
will discuss - a black hole. It does {\it not} contribute to the
Rademacher sum because the restriction on the sum in \morphys\ is
given by a strict inequality. This is also the value of $J_0$
 at which the number of chiral primaries
is  a maximum.  For an antichiral primary the
spectral flow invariant is $-(J_0 + k)^2$ and a similar discussion
applies.

\ifig\symnsstates{A more symmetric presentation of the NS particle
region.  By integer spectral flow \spectfl\ with $\theta=-1$ we
can map the upper region in \nsstates\ to the left, producing  a
domain symmetric about the $L_0$
axis.}{\epsfxsize3.0in\epsfbox{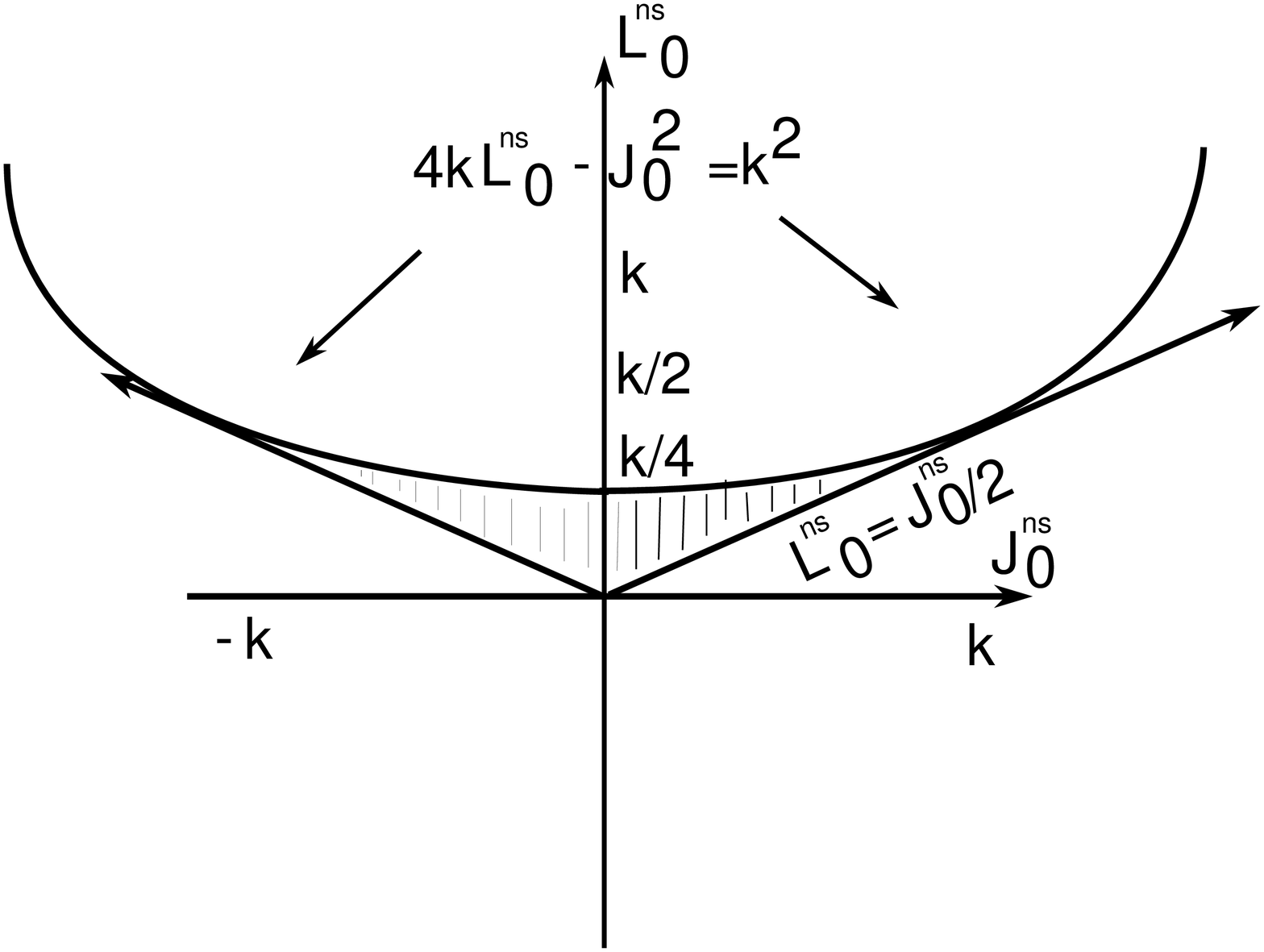}}

\subsubsec{Supergravity interpretation}

 Now let us turn to the
supergravity solutions which should contribute in the ADS/CFT
correspondence. The Jacobi-Rademacher series involves a sum over
quantum numbers of the $SU(2)_L$ part of the $SU(2\vert 1,1)$ AdS
algebra. In supergravity these quantum numbers are associated with
the $SU(2)_L$ Kaluza-Klein gauge theory from isometries of
$S^3=SU(2)$ in the 6d geometry $AdS_3 \times S^3$. Therefore, we
should study supergravity solutions which are locally $AdS_3
\times S^3$ and associated to spinning black holes. These have
been discussed by Cvetic and Larsen \cvl, and their solution may
be summarized as follows. The Lorentz-signature geometry is given
by a metric of the form
 \eqn\cvlmet{ ds^2 = ds^2_{BTZ} + A^a \otimes A^b
(K^a , K^b) + A^a \otimes K_a + K_a \otimes A^a +
ds^2_{S^3} } where $a$ is an adjoint $so(4)$ index and $K^a$ are
$so(4)$ Killing vectors for the round metric $ds^2_{S^3}$ and
$K_a$ are the dual one-forms. Explicitly, we write the metric on
$S^3$ as
\eqn\cvlmetii{ ds^2_{S^3} = d\phi^2 + d\psi^2 + d \theta^2 + 2
\cos\theta d\phi d \psi}
in terms of the standard
 \foot{ Warning:
 \cvl\ uses the same notation for angles which are not
 Euler angles. We have $\psi^{C.L.} = \half (\psi + \phi),
 \phi^{C.L.} = \half (\psi - \phi), \theta^{C.L.} = \half
 \theta$.}
 Euler angle parametrization of
 $g\in SU(2)$:
 \eqn\eulerang{
 g(\phi,\theta,\psi):= \exp[i \half \phi \sigma^3] \exp[i \half \theta
 \sigma^1] \exp[i \half \psi \sigma^3], \qquad  0 \leq \psi\leq 4\pi,
 0 \leq \phi\leq 2 \pi, 0 \leq \theta\leq \pi.
 }
In these coordinates,   $K_{3,L} = d \psi + \cos \theta d\phi,
K_{3,R}= d \phi + \cos\theta d \psi$, are dual to $K^{3,L} = {\p
\over \p \psi}, K^{3,R} = {\p \over \p \phi}$, respectively.
$ds^2_{BTZ} $ is the standard Lorentz-signature BTZ black hole
with coordinates $t,r, \phi_b$, $r\geq r_+$, $\phi_{b} \ \sim
\phi_b + 2 \pi$. In these coordinates the Kaluza-Klein gauge
fields are given by
\eqn\cvlgflds{ \eqalign{ A_L^a & = - 4 \delta^{a,3}{j_L \over k} (dt +
d \phi_b ) \cr  A_R^a & = - 4 \delta^{a,3}{j_R \over k} (dt -  d \phi_b
) \cr} }

Notice that $A_L, A_R$ are left- and right-chiral flat gauge
fields on the family of tori at fixed $r$ foliating the BTZ black
hole. They have nontrivial monodromy around the spatial cycle but,
because there is a hole in space, there is no singularity in the
geometry. Moreover, they can be removed by an improper  gauge transformation
on the $S^3=SU(2)$ coordinate $g(\phi, \theta, \psi)$ by:
\eqn\ggetmn{ g \rightarrow \tilde g:= e^{-i {j_L \over k} (t  + \phi_b)
\sigma^3} g
 e^{-i {j_R \over k } (t  - \phi_b) \sigma^3}
}
This is not a good gauge transformation in general since it is not
periodic in $\phi_b$.
The spins  $j_L \pm j_R$ are integrally quantized $so(4)$ spins in
the quantum theory.
 This
transformation brings the metric to the standard $AdS_3 \times
S^3$ form and allows us to complete the solution of the 6d (0,2)
supergravity by writing the $3$-form fieldstrength (for one of the
5 self-dual $H$ fields of the $(0,2)$ supergravity from IIB on K3)
as:
\eqn\hai{ H =  dt\wedge  dr\wedge  r d \phi_b +  {1 \over
12 \pi} {\Tr}(\tilde g^{-1} d \tilde g)^3 }
Note that when expanded using \ggetmn\ the $H$ field will contain
terms proportional to $dt \pm  d \phi_b$.

In the AdS/CFT correspondence the above geometries correspond to
semiclassical states with $J_0 = 2j_L, \tilde J_0 = 2 j_R$. In
\cvl\ it is shown that the cosmic censorship bound translates into
the inequalities
\eqn\ccship{ 4k L_0 - J_0^2 \geq 0 \qquad 4k \tilde L_0 - \tilde
J_0^2\geq 0 }
{\it We thus interpret the lattice points $(\mu,m)$ in the
particle region of \nsstates\ as quantum numbers of   particles
which are not sufficiently massive to form black holes. } The sum
over the particle region in the Jacobi-Rademacher formula will
thus involve a sum over these particles.

Now let us turn to the analytic continuation of the Cvetic-Larsen
solutions. It is important to bear in mind that these represent
saddle points in a thermal ensemble, and are not semiclassical
descriptions of states with well-defined mass and spin in a
quantum gravity Hilbert space.  The Euclidean continuation of the
BTZ geometry is $t \rightarrow i t_E, \phi_b \rightarrow \phi_b,
r\rightarrow r, r_+ \rightarrow  \tau_2, r_- \rightarrow i
 \tau_1$, producing the geometry \eucbtz. We also continue
$j_L \rightarrow \omega_L, j_R \rightarrow \omega_R$ where
$\omega_L, \omega_R$ are complex spin fugacities. The resulting
Euclidean solution has several interesting features.

First, the gauge fields are complex. Moreover, after Euclidean
continuation the time coordinate $t_E$ becomes periodic and the
circle of time is {\it contractible} in the solid torus topology
of \eucbtz. Thus the flat gauge fields have Wilson-line
singularities at the ``center'' of the solid torus $r =
\tau_2$, namely $F_L^a \sim \delta^{a,3} \omega_L
\delta^{(2)}(t_E, \phi)$. Similarly, $H$ picks up a singularity,
indicating the presence of a string at the Wilson line. The flat
gauge fields
\eqn\cvlgflds{
\eqalign{ A_L^3 & = \omega_L (d \phi + i dt_E) \cr A_R^3 & =
\omega_R (d \phi - i dt_E) \cr} }
are precisely of the right form to agree  with the boundary
conditions on the Euclidean path integral \aleftgf\arightgf\
appropriate for evaluation of the $SU(2)_L \times SU(2)_R$
Chern-Simons path integral.

 Thus, we propose that the Cvetic-Larsen solutions are saddle
point approximations to the geometries that contribute in the AdS
path integral dual to the elliptic genus. We incorporate the sum
over lattice points $(m,\ell)$ in the particle regions of
\rstates, \nsstates, \symnsstates\ as follows.

In the full theory there are (presumably smooth) 6d or (at shorter
distances) 10d geometries which involve particles propagating
along  ``worldlines'' $\gamma$. At  long distances the metric  is
locally   $AdS_3$ and best described by a Chern-Simons gauge field
${\bf A}$ of an $SU(2\vert 1,1)_L \times SU(2\vert 1,1)_R$
Chern-Simons theory.   The particles with worldline $\gamma$ are
described by the Wilson line ${\Tr} P\exp[\int_\gamma {\bf A} ] $.
It would be very interesting to find explicit smooth solutions of
the 6d $(0,2)$ supergravity equations corresponding to these
particles. Consistency of this picture demands that we should only
sum over particles which themselves {\it do not} form black holes.
Accordingly, we should include Wilson lines for the
representations corresponding to the quantum numbers $(m,\ell)$
which are   lattice points  in the particle region of \nsstates.

\bigskip
\noindent {\bf Remarks}

\item{1.} Notice that when we perform the \holographic transform
we omit the state with $(m,\ell)=(0,0)$. This state is
special because it is the unique state that is both a black hole and
a chiral primary. From this point of view it seems natural that one
should remove  it from the partition function so that
we get a simple expresion such as \introvii \morphys .


\item{2.} As we have stressed above, the interpretation of the
geometries as having an insertion of a Wilson line   resolves a
paradox regarding the continuation into three dimensions of the
odd spin structure on the rightmovers required for computation of
the elliptic genus.

\item{3.} The physical interpretation of the ``worldline'' $\gamma$
depends on the physical interpretation of the solid torus
geometry. If the geometry is that of  Euclidean thermal AdS then
$\gamma$ is indeed a worldline. However, in a black hole geometry,
$\gamma$ is a {\it spatial} cycle at a fixed Euclidean time. Thus
the Wilson line is associated with the virtual particles
associated with Hawking radiation from the black hole.
Indeed, in this interpretation of the spatial Wilson lines they
represent a sequence of pair creation processes of the Hawking particles
that surround the black hole like a virtual cloud. The contribution of
these particles is included through a multi-particle generalization of
the Schwinger calculation, and, just as for the original Schwinger
calculation, give us information about the probability of pair creation
in the gravitational field of the black hole. The fact that the Wilson
lines are non-contractible makes clear that the processes involve
particle-antiparticle pairs that, in the Euclidean world, go once or
more times around the black hole before they again annihilate. The
quantum mechanical probabilty of such a process is exponentially
suppressed with an exponent that is proportional to the length of this
euclidean path, which for the BTZ-black hole is (an integer multiple of)
$\rm{Im}\left( -1/\tau \right) .$ \ In this interpretation
the black hole Farey tail
represents the trace of the density matrix of Hawking
particles outside a black hole. Only, just as for the thermal AdS,
the trace is again truncated to the subset of those ensembles of Hawking
particles which themselves do not form a black hole.

\subsubsec{Remark on the particle degeneracies $c(m,\ell)$.}

Notice that the only remnant of the fact that we have compactified
the underlying microscopic superstring theory on a $K3$ surface
$X$ is in the degeneracies $c(4km-\ell^2;{\rm Sym}^k X)$. The
question arises as to whether these degeneracies themselves can be
deduced purely from supergravity. In \deboer\ it is shown that,
at least for the lattice points of \symnsstates\ with
$L_0^{NS}\leq {k\over 4}$, the degeneracies can indeed be obtained
from Kaluza-Klein reduction of $IIB$ supergravity on $AdS_3 \times
S^3 \times X$. Extending this result to the
full region of \symnsstates\ is an interesting open problem.

It is also worth stressing that the degeneracies
 $c_\ell(4km-\ell^2;{\rm Sym}^k X)$ can be quite large.  They can be
extracted from the formula \dmvv
\eqn\bproduct{ \sum_{k=0}^\infty p^k \chi({\rm Sym}^k X; q,y) =
\prod_{n>0, m\geq 0, r} {1 \over  (1-p^n q^m y^r)^{c(nm,r)}} .}
In particular note that
  by taking $q \rightarrow 0 $ we get the
well-known result of \gottschsoergel
\eqn\rrdegen{
\sum_{k=0}^\infty \sum_{\ell=0}^{2k}
 \chi_{\ell}({\rm Sym}^k X) p^k y^{\ell-k}
= \prod_{n=1}^\infty { 1 \over  (1 - p^n y^{-1})^2 (1-p^n)^{20} (1
- p^n y^{+1})^2} }
%
where $\chi_{\ell}(M):=\sum_s (-1)^{s+\ell} h^{s,\ell}(M)$ for a
manifold $M$. Using \rrdegen\
 we can obtain degeneracies at $m=0$.
 As we scan
 $\mu$ from $-k$ to $0$ at $m=0$  the degeneracies increase from
 $c(-k^2;{\rm Sym}^k X) = k+1$ to   $\sim \exp[ 4 \pi \sqrt{k} ]$ near
 $\mu \cong 0$.  Expressed in   terms of the geometry
 of the D1D5 system, this degeneracy near $\mu \cong 0$ is
 of order $\sim \exp[ 4 \pi
\sqrt{r_1 r_5}/g_{str} ]$, in the notation of \stringexclusion. In
particular, there is a nonperturbatively large ``ground state''
degeneracy.

\subsec{Interpreting the gravitational factor}

We now propose that the factor
\eqn\grvtyfctr{
\bigl(  c \tau + d  \bigr)^{-3}
\exp\biggl[2\pi i \bigl(  m -{ \mu^2\over
4k}\bigr) {a \tau + b \over  c \tau + d} \biggr]
}
is the contribution of $SL(2,\IC)$ Chern-Simons
gravity.

Note first that the contribution of the $SU(2\vert 1,1)$
Chern-Simons path integral defines a wavefunction on the universal
elliptic curve, parametrized by $(\tau,\omega)$. This should be a
modular invariant half-density $Z(\tau,\omega) d\omega \wedge d
\tau$, so the modular transformation law given by $\bigl( c \tau +
d \bigr)^{-3}$ is just right.

We now interpret the remaining exponential
as the holomorphic part of
the Euclidean gravitational action
\eqn\gravactp{
S = {1 \over  16 \pi G}
\biggl[ \int_{M}  d^3x \sqrt{-g} (\CR - 2 \Lambda) +
2 \int_{\p M} K \biggr]
  }
for a BTZ black hole.

A straightforward computation shows the following. Paying due
attention to the identification of space and time with $a$,$b$-cycles
we conclude that if we define the gravitational action of
$BTZ(\tau)$ (where $\tau$ is defined with $a$-cycle = contractible
cycle) relative to an AdS background
\foot{The action \gravactp\ is infinite, even including the
boundary term. Thus one actually computes differences of actions
for pairs of geometries with diffeomorphic asymptotics.}
 then in fact the gravitational action is:
\eqn\upsdown{ {\pi \ell\over 4 G} \Im (-1/\tau). }
Using $\ell/4G = k $, the    generalization of this result to the
$SL(2,\IZ)$ family gives the gravitational action
\eqn\gengrvact{
\exp[ 2 \pi i \tau' h  - 2 \pi i \bar \tau' \tilde h] }
where
$\tau' = (a \tau + b)/(c \tau + d)$. In \cvl\ Cveti\'c and Larsen
show that extremal rotating black holes have  $\tilde h = 0 $ and
$h = m- \mu^2/4k $ so we obtain the expression in \morphys.

\subsec{Interpreting the $SU(2)$ factor}

Finally we interpret the factor
\eqn\sutwofct{
\exp[- 2\pi i k {c \omega^2 \over  c \tau +d}]
\Theta^+_{\mu,k}({\omega\over  c \tau +d} ,
{a \tau + b \over  c \tau + d}  )
}
in \morphys\ as the contribution of the
$SU(2)_L$ Chern-Simons theory to the
path integral.

The functions $\Theta^+_{\mu,k}(z,\tau)$ are closely related to
the wavefunctions for $SU(2)$ Chern-Simons theory. It follows from
the general reasoning of \jones\ that the $SU(2)$ level $k$
Chern-Simons path integral on a solid torus may be expanded (as a
function of $z$) in terms of these functions. This was carried out
more explicitly in   \emss. One basis for the level $k$ $SU(2)$
Chern-Simons wavefunctions is given by the characters of affine
Lie algebras: \eqn\cswvfn{ \eqalign{ \Psi^{CS}_{\mu,k}(z,\tau) & =
\exp\biggl( \pi k {z^2 \over \Im \tau}\biggr)
{\Theta_{\mu+1,k+2}(z,\tau) - \Theta_{-\mu-1,k+2}(z,\tau) \over
\Theta_{1,2}(z,\tau) - \Theta_{-1,2}(z,\tau) } \cr} } This basis
diagonalizes the Verlinde operators and is naturally associated
with the path integral on the solid torus with a Wilson line in
the spin $j = \mu/2$ representation labeling the $b$-cycle.

The space of wavefunctions (as functions of $z$) spanned by
\cswvfn\ is the same as the space spanned by even level $k$ theta
functions because of the identity
\foot{The ``parafermion'' terms $c_{\mu,\mu'}(\tau)$ may
themselves be written in terms of higher level thetanullwherte.}
\eqn\parafermi{
{\Theta_{\mu+1,k+2}(z,\tau) - \Theta_{-\mu-1,k+2}(z,\tau) \over
\Theta_{1,2}(z,\tau) - \Theta_{-1,2}(z,\tau) }
  = \sum_{\mu'=0}^k c_{\mu,\mu'}(\tau)
\Theta_{\mu',k}^+ (z,\tau) }

Now
consider the exponential prefactor in \sutwofct.
Take the Chern-Simons wavefunction
\eqn\consdi{
 \exp\biggl( \pi k {\omega^2 \over \Im \tau}\biggr)
\Theta^+_{\mu ,k}(\omega,\tau) }
and substitute
 \eqn\thensubs{ \omega \rightarrow {\omega \over c
\tau + d} \qquad \tau \rightarrow {a \tau + b \over c \tau + d} .}
In the full elliptic genus only rightmoving BPS states contribute.
Therefore we can take $\tau, \bar \tau$ to be independent and only
the term surviving the limit   $\bar \tau \rightarrow -i \infty$
can contribute to the elliptic genus. This limit gives   the
expression in \morphys:
\eqn\holowvfn{
 \exp[- 2\pi i k {c \omega^2 \over  c \tau + d}]
 \Theta^+_{\mu,k}({\omega \over  c \tau + d} ,
{a \tau + b \over  c \tau + d}  ) . }

Note that the basis of $SU(2)$
Chern-Simons wavefunctions $\Theta^+$ is preferred  over the
 character basis \cswvfn. The character
basis is usually thought of as the
preferred basis because it diagonalizes the
Verlinde operators. The function
$\Theta^+$ sums states at definite values
of $J^3_{0,L}$ in the current algebra, while \cswvfn\ sums states
at definite values of the Casimir, $j(j+1)$. From the view of
the sum over geometries, the $\Theta^+$
basis is preferred because the geometries are at definite values of
$J^3_{0,L}$.

\subsec{Comparison to previous approaches to
the quantum gravity partition function and the
Hartle-Hawking wavefunction}

It is interesting to compare the above
result for the supergravity
partition function on the
solid torus with the results of more
traditional approaches to quantum
gravity. The standard approaches,
in the context of 3-dimensional
gravity are described  in
\carlipbook.

The elliptic genus is a sum over
states.
By the AdS/CFT correspondendence
these states can be identified with states
in the supergravity theory on $AdS$, with
the $AdS$ time defining Hamiltonian
evolution. In particular, the quantum gravity
has a well-defined Hilbert space!

When time is made Euclidean and periodic it is traditional to replace
the sum over states by a sum over Euclidean geometries. What has never
been very clear is what class of geometries and topologies one should
sum over. In the case of Euclidean geometries bounding a torus this
question has been explicitly addressed by Carlip in \carlipsum. In the
case of a negative cosmological constant there are finite volume
hyperbolic geometries which bound the torus (see, e.g., \elstrodt). These
are hyperbolic three-manifolds with a cusp, i.e.\
a boundary at $r=\infty$ where the metric behaves as
\eqn\cusp{ ds^2 \sim { 1\over r^2} (dr^2+ g_{ij} dx^i dx^j),}
This is the opposite behaviour of the one considered in the AdS/CFT
correspondence as induced by the D-branes.\foot{It would be very
interesting to find the appropriate interpretation (if any) of these
cusps in the AdS/CFT correspondence.}
 There are no uniqueness theorems
for manifolds with this boundary behaviour.
The entropy of these geometries overwhelms the action, making the
Hartle-Hawking wavefunction ill-defined. In the AdS/CFT formulation of
quantum gravity these geometries are eliminated by the boundary
condition. The gravitational action of the infinite volume hyperbolic
geometry must be regulated, but once this is done the sum over
topologies (i.e. the sum over $(c,d)$) is well-defined and convergent.

\subsec{Puzzles }

It will be clear, to the thoughtful reader, that the physical
interpretation we have offered for the formula \morphys\ is not
complete. We record here several puzzles  raised by
the above discussion.

\item{1.} It would be nice to understand more clearly the
hypothetical smooth six-dimensional geometries corresponding to
``adding particles to black holes.'' Related to this, it would be
nice to go beyond the asymptotic topological field theory and
understand more fully the $SU(2)_L \times SU(2)_R$ gauge theory
that arises in $(0,2)$ supergravity on $AdS_3 \times S^3$. For
some preliminary remarks see \degersezgin.

\item{2.} The procedure of taking $\bar \tau \rightarrow - i \infty$ in
\holowvfn\ is {\it ad hoc}. Clearing up this point requires a careful
discussion of the inner product of wavefunctions in the  coherent
state quantization of the full $SU(2\vert 1,1)$ Chern-Simons
theory.

\item{3.} It would also be interesting to understand more precisely
the physical meaning of the \holographic transform. From the fact
that the partition function becomes a wave function it seems reasonable
to think that it  corresponds to extracting some singleton degrees
of freedom that live at the boundary of $AdS$. Another hint is that
the \holographic  transform is just Serre duality, from the
mathematical perspective. This is reminiscent of the fact that
in the AdS/CFT correspondence supergravity modes and CFT operators
are not equal, but rather in duality.

\newsec{Large $k$ phase transitions}

In this section we will derive the
$SL(2,\IZ)$ invariant phase structure at
large $k$ as a function of $\tau$.
\foot{The phase structure of the D1D5 system
has been
discussed in a different way in \martsahak.
 }

We begin with the form \bozo\polarpart\ of the Jacobi-Rademacher
expansion. We go to the NS sector by spectral flow:
$Z_{NS,R}(\tau,0) = (-1)^k e^{2\pi i \tau k/4} Z_{R,R}(\tau,\omega=\tau/2)$.
The net result for $Z_{NS,R}(\tau,0)$  is
\eqn\netresult{  e^{i\pi k/2} \sum_{(c,d)=1, c\geq 0} ( c \tau +
d)^{-3} e^{- 2\pi i k {c (\tau/2)^2 \over  (c \tau +d)} }
 \sum_{4km-\ell^2<0} \tilde c_\ell(4km -\ell^2)
e^{2\pi i m {a \tau + b \over  c \tau + d}  + 2 \pi i \ell
{\tau/2 \over  c \tau + d} }
}

We will estimate the magnitude of the various terms in the sum
\netresult. In order to do this we   need the following identities
(valid for $c\not=0$):
 \eqn\simplestf{ \eqalign{ {a \tau + b
\over c \tau + d} & = {a \over  c} + {- 1 \over  c (c \tau +d)}
\cr {\tau/2 \over c \tau + d} & = {1 \over 2 c} + {d \over 2} {- 1
\over  c (c \tau +d)} \cr
 {c (\tau/2)^2
\over  (c \tau +d)} & = { \tau \over 4} - {d \over  4 c} - ({d \over  2})^2
 {- 1 \over  c (c \tau +d)} \cr}
}
  Using these identities one
can evaluate the absolute norm of the terms in
the sum \netresult. We find the norm:
\eqn\normtrms{
{1 \over  \vert c \tau + d \vert^3} \vert \tilde c(4km-\ell^2) \vert
\exp\biggl[  - 2 \pi
\Im ({a\tau + b \over c \tau + d} ) \biggl( k ({d \over  2})^2 + m +
\ell {d \over  2} \biggr) \biggr]
}
 We will
show that the sum (on $m,\ell$)  is bounded by a constant
for large values of $c,d$. Then the sum over
$(c,d)$ is absolutely convergent, because of
the factor $( c \tau + d)^{-3}$.  Therefore,
to study the large $k$ limit we can study the
terms for fixed $(c,d)$ separately. For fixed
$(c,d)$ we now analyze which terms in the
sum over $(m,\ell)$ dominate in the large $k$
limit. The sum over
$(m,\ell)$ is over lattice points between
the parabolas $4km - \ell^2 = 0$ and $4km-\ell^2 = -k^2$.
Since $\Im ({a \tau + b \over  c \tau + d})>0$,
  we minimize as a function of $m$ for
 $m = {\ell^2 \over  4k} -{k \over  4}$.
Next we minimize with respect to $\ell$. The minimum is taken at
$\ell_* = - k d$, $m_* = k (d^2-1)/4$. (In order to guarantee that
this is an integer we take the limit $k \rightarrow \infty$ with
$k$ divisible by $4$.) Noting that $\vert \tilde c(4km-\ell^2)
\vert$ is bounded by a constant depending only on $k$, we conclude
that  the dominant term for fixed $(c,d)$ has magnitude:
\eqn\dominant{ \exp\biggl[ 2 \pi {k \over  4} {\im \tau \over
\vert c \tau + d \vert^2}\biggr] } and all other terms in the sum
on $m,\ell$ are exponentially smaller. This statement also holds
for the case $(c=1,d=0)$. If we now consider the sum over $(c,d)$
we see that in each region of an $SL(2,\IZ)$ invariant
 tesselation of the  upper half plane (corresponding
to the keyhole region and its modular images) there is a unique
$(c,d)$ which dominates. The reason is that the keyhole region
has the property that the modular image of any point
 $\tau\in \CF$ has an imaginary part $\im \tau' = \im \tau/\vert c \tau + d\vert^2
\leq \im \tau$. Thus the phase transitions are located at the
boundary of the region $\Gamma_\infty\cdot \CF$ and its images
under $SL(2,\IZ)$.

Thus we conclude  that $\log Z_{NS}$ is a piecewise continuous
function with discontinuous derivatives across the boundaries of
the standard $SL(2,\IZ)$ invariant tesselation of the upper half
plane: There are first order phase transitions across the
boundaries of the $SL(2,\IZ)$ invariant tesselation.

\ifig\freeen{The free energy computed from
$Z_{NS}$ in the $k \rightarrow \infty$ limit,
as a function of inverse temperature $\beta$. The dashed line indicates
a typical  finite $k$ result.   }
{\epsfxsize3.0in\epsfbox{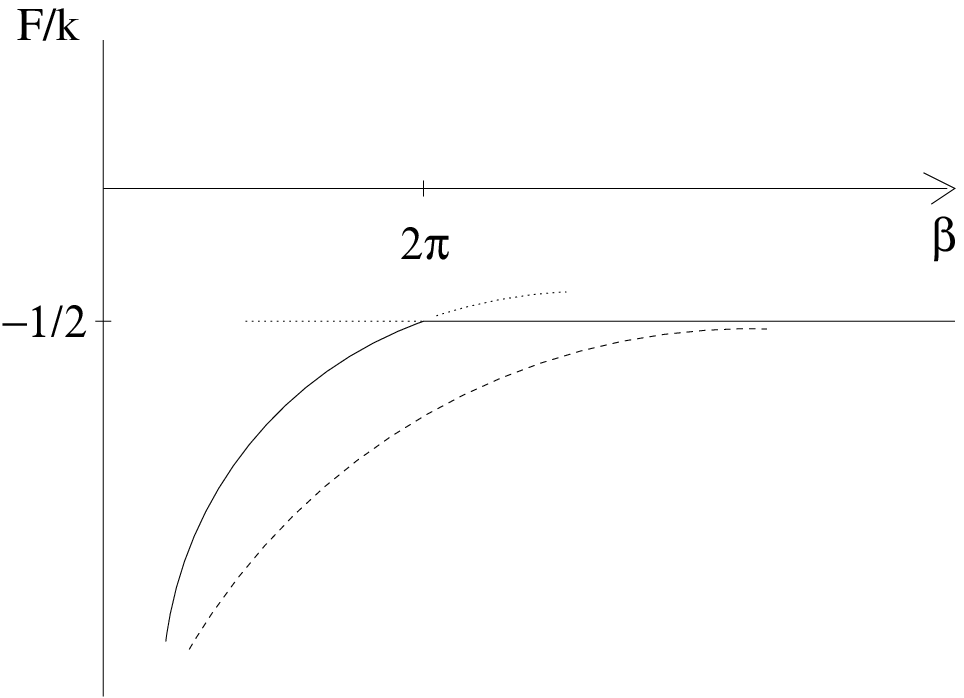}}

In order to understand the physics of the phase transitions more
clearly let us  focus on the phase transition for  along the
imaginary $\tau$ axis, $\Re(\tau)=0$. There is a phase transition
as $\Im \tau$ crosses from $1-\epsilon$ to $1+ \epsilon$, and
the derivative of the
 free energy exhibits a discontinuity as shown in \freeen. The
free energy is defined to be $F:=-{1 \over \beta} \log
Z_{NSNS}(\tau={i \beta\over  2\pi} )$. At low temperatures the
contribution of the leftmovers to the free energy is:
\eqn\lowfree{
 F = -{ c \over 24} = -{k\over 4} }
while at high temperatures the contribution
of the leftmovers to the free
energy is
\eqn\highfree{
F = - {k \over 4 } { 4 \pi^2  \over \beta^2 }
. }
We have done the calculation for the elliptic
genus with $(NS,R)$ boundary conditions.
If we want both left and rightmovers to have NS boundary
conditions the formulae \lowfree\highfree\ get multiplied
by 2.

We now give  an heuristic argument which explains the physical
nature of this phase transition. Consider $IIB$ theory on $AdS_3
\times S^3 \times X$ where $X$ is a $K3$ surface (or a torus $T^4$).
 At the orbifold point, a CFT with
target space ${\rm Sym}^k(X)$ is equivalent to a gas of strings
with total winding number $k$ moving on $X$ \dmvv.  We are
interested in putting this conformal field theory on a circle with
antiperiodic (NS) boundary conditions for the fermions. There is a
unique ground state where  all strings are  singly wound  and in
their ground states on $X$. The lowest energy state is when they
are singly wound because the twist field that multiply winds them
has positive conformal weight, of order $\Delta \sim w/4$.
The energy of this NS vacuum  state is \lowfree.
We can excite the system by  putting oscillations on the strings
or by multiply winding them. Since we have to symmetrize the state
we could think of these excitations as creating  second quantized
string states in a multiple string hilbert space \dmvv. The number
of such states is independent of $k$ at low energies $ E \ll k$.
The energies of these states are of order one. They will not
contribute very much to the free energy at temperatures of order
one. Another set of states can be obtained by
 multiply winding the
strings. If we multiply wind a string $w$ times we have to
supply an energy of the order of $w/4$ but we decrease the energy gap of
the
system which is now of the order of $1/w$
\maldasusskind.
 \foot{The analysis in \maldasusskind\
 considered the system
in the Ramond sector  where
 strings can be multiply wound  at no energy cost.}
 Then we can have an entropy of
the order of $S = 2 \pi^2 w/\beta$ coming from the oscillations of the
string. Notice that we can apply the large temperature approximation
for calculating the entropy for large $w$, when the gap is very small.
Taking into account the energy of the oscillations and the
energy necessary to multiply wind it we see that the contribution of
these states to the free energy is $F = E-TS \sim w/4(1 -4 \pi^2/\beta^2)$ so
that it becomes convenient to produce them above a critical temperature,
 $\beta < 2\pi$.
 The maximum winding number is $k$ and we see that then
the free energy becomes \highfree\ (accounting for the factor of
two as noted above.)  On the other hand, approaching the phase
transition from the low temperature side we see a Hagedorn density
of string states, with a Hagedorn temperature
 $ T =1/(2 \pi)$ (see below),
so as long as the temperature is smaller than this, the
 contribution of these states to the free energy is finite and independent
of $k$ and therefore subleading compared to \lowfree .
This Hagedorn density of states appears only at the orbifold point and it
is not seen in supergravity. It appears from the supergravity analysis that
most of these states get large masses.

{}From the supergravity point of view this is the usual Hawking-Page
transition between thermal AdS and a black hole background
as descibed in \wittenads .
Actually we need to be more precise since we are considering only the
left movers and we are considering the R ground states on the right. If we
start in the (NS,NS) vacuum, then the R ground states on the right correspond
to chiral primaries. So we need to understand which black holes are
chiral primaries on the right.
%
 Here again we use the
  cosmic censorship bound of
   \breckenridge\cveticyoum\cvl.
As we have discussed, this bound implies  $\tilde  L_0 \geq \tilde
J_0^2/4k$ with  a similiar bound for $L_0 $ and $J_0$. This bound
is in general stronger than the chiral primary bound, except for
$\tilde J_0 = k$ when they coincide (see \nsstates.).
So the black hole will have these right-moving quantum numbers.
The left
moving part of this black hole has essentially the same free
energy as the BTZ solution (we take $J_L =0$ in the absence of
chemical potentials for $J_L$). So that the free energy is
$F^{sugra} = - {1 \over 4} {4 \pi^2 \over \beta^2} $.

The smallest value of $L_0$ for a black hole consistent with
cosmic censorship is $L_0 = k/4$.   This naturally ``explains''
why de Boer \deboer\  found agreement with supergravity up to this
stage; at this point black holes start contributing to the
elliptic genus.
Beyond this point we only seem to find agreement in the asymptotic form
of the sugra and CFT elliptic genus.
 In fact, beyond this point the CFT form of the
elliptic genus and the gravity form for it continue to agree to
leading order in $k$, they both show an exponentially large
(Hagedorn) number of states $e^{2 \pi \sqrt{k h }}$ where $h = L_0
  $ is the energy in the NS sector relative to the NS ground
state.
Notice that our results indicate that if we include all black hole
contributions and all particles around them that do not form black
holes then we  also find agreement.

Similarly notice that the extremal rotating black hole with $\half
J_0 =\half  \tilde J_0 =L_0 =\tilde  L_0 = k/2$ is a (chiral,
chiral) primary. This   gives a reason  for expecting that
disagreement between supergravity and CFT spectra of (chiral,
chiral)   appears when $J_0 = \tilde J_0 =k $. Indeed this is the
point where the ``exclusion principle'' becomes operational
\stringexclusion\deboer.

{\bf Remarks}

\item{1.} There is an interesting analogy
between the phase transitions discussed here and those in the
four-dimensional $U(N)$ case discussed by Witten \wittenads. The
analogy is that the permutation symmetry is analogous to the gauge
group, i.e. $S_k$ is analogous to $U(N)$. In the low temperature
phase all oscillations of different strings have to be symmetrized
(i.e., made   gauge invariant) and this reduces the number of
independent excitations, which becomes independent of $k$. On the
other hand in the high temperature regime, it is as if all strings
are distinct and one can put independent excitations on each of
the strings. So in this high temperature phase the permutation
symmetry is ``deconfined.'' Of course what is making this possible
is the multiple winding since the whole configuration should be
gauge invariant.

\item{2.}   $SL(2,\IZ)$
invariant phase diagrams have appeared before in different
contexts. In four-dimensional abelian lattice gauge theory one
finds such transitions as a function of $\tau= \theta + i/e^2$
\ref\cardy{J. Cardy, Nucl. Phys. {\bf B205}(1982)17}. Also, in the
dissipative Hofstadter model one finds an $SL(2,\IZ)$ invariant
phase diagram as a function of $\tau = B+ i \eta$ where $B$ is a
magnetic field and $\eta$ is the dissipative parameter of the
Caldeira-Leggett model. See \ref\callanfreed{C. Callan and D.
Freed, Nucl Phys. {\bf B374} (1992) 543; Nucl. Phys. {\bf B392}
(1993) 551}. The phase boundaries in these models are
Ford circles (see appendix B), which are modular equivalent
to $\Im \tau=1$. In our example, as we have explained,
the phase boundaries are at the boundary of the region $\Gamma_\infty\cdot \CF$
and its images under $SL(2,\IZ)$.

 \newsec{Conclusions and Future Directions}

In this paper we have applied some techniques and results of analytic
number theory to the study of supersymmetric black holes. This has lead
to a formula \morphys\ for the elliptic genus. Using this formula in
\netresult\ we were able to derive an $SL(2,\IZ)$ invariant phase
diagram. We have also offered some physical interpretations of these
formulae. It should be stressed that if it turns out that there are
flaws in these interpretations this would not invalidate the basic
formula \morphys, nor the derivation of the phase diagram in section
six. There are several interesting avenues for further research and
possible applications of these results. Among them are:

1. The Jacobi-Rademacher series gives a useful way of controlling
subleading terms in the exact entropy formula for black hole degeneracy
provided by the elliptic genus. For this reason the result \morphys\
might help in proving or disproving some of the conjectures of
\millermoore\ relating black hole entropy to some issues in analytic
number theory.

2. It is interesting to compare with Witten's discussion of the
partition function of $d=4$ $U(N)$ gauge theory via the AdS/CFT
correspondence in \wittenads. There are only two obvious ways to fill in
$S^{n-1}\times S^1$ topologically for $n>2$, but in our case, with
$n=2$, there are infinitely many ways. Note that the right-moving odd
spin structure of the elliptic genus at first sight suggests that we
cannot fill in the torus with a solid torus at all. This did not in fact
kill the sum over instantons because of the the Wilson line defects.
Thus, this example raises the question of whether similar things might
also happen in higher dimensions, i.e., whether there might be other
terms in the Euclidean partition sum besides the geometries $X_1, X_2$
considered in the calculation of Witten.

Another lesson for higher dimensional calculations is that other
boundary geometries with nontrivial diffeomorphism groups (such as
$T^4$) will probably lead to interesting infinite sums over instanton
contributions. It would be quite interesting to reproduce, for example,
the formulae of Vafa and Witten from the ADS/CFT correspondence
\vafawitten.

3. Notice that  the exact result for the elliptic genus
 turns out to depend very little on most of the data one needs
to define the full  partition function on $AdS_3 \times S^3$. For example, we
did not need to specify the boundary conditions for the
massive fields.
In general the full partition function will depend on these boundary
conditions, it is only for a
 semi-topological quantity like the elliptic genus that these do not enter.

4. There are several variations on the above results which would
be interesting to investigate. It should be straightforward to
extend the above discussion  for $K3$ to the case of $T^4$. This
will involve a $\tilde J_0^2$ insertion in the path integral, as
in \mms. One can also ask about higher genus partition functions,
as well as about extensions to nonholomorphic quantities such as
the NS-NS partition function at $\omega= \tilde \omega=0$.

5. It would be interesting to see if similar formulae apply to the
other AdS $(p,q)$ supergravities of \townsend.

\bigskip
\centerline{\bf Acknowledgments }\nobreak
\bigskip

This paper has had a somewhat extended gestation. For two of us (JM and
GM) it began in collaboration with A. Strominger at Aspen in August
1997. We thank A. Strominger for collaboration at that time and for
several important discussions since then. We also thank S. Miller for
collaboration on the mathematical status of the Rademacher expansion for
the case of $w=0$ and for comments on the manuscript.
 We would also like to thank R. Borcherds, J. de Boer,
M. Flohr, V. Gritsenko, J. Harvey, F. Larsen, E. Martinec, C. McMullen,
N. Read, N. Seiberg, E. Witten, and D. Zagier for discussions and
remarks. JM and GM would like to acknowledge the hospitality of the
Aspen Center for Physics,  the Amsterdam Summer Workshop on String
Theory and Black Holes and the Institute for Advanced Study.
 RD would like to thank the Yale Physics
Department and the Institute for Theoretical Physics, Santa Barbara for
hospitality. GM was supported by DOE grant DE-FG02-92ER40704 at Yale,
and is now supported by DOE grant  DE-FG02-96ER40949. JM is
supported  in part by DOE grant  DE-FG02-91ER40654 and by the Sloan and
Packard foundations.

Version 3 of this paper was stimulated by an important observation
of Don Zagier, as explained in the introduction.

\appendix{A}{Notation $\&$  Conventions}

\medskip
$A, \tilde A$ \qquad $SU(2) \times SU(2)$ gauge fields from
KK reduction of 6d gravity on $AdS_3 \times S^3$.

\medskip
$\beta$ \qquad Inverse temperature. Also, the analytic
continuation of $\beta$. It might be complex.

\medskip
$c$ \qquad A central charge. We use $c=6k$ instead
to avoid confusion with integers in $SL(2,Z)$ matrices.

\medskip
$c,d$ \qquad  Relatively prime integers. Often entries in an
$SL(2,\IZ)$ matrix $\gamma$.

\medskip
$\hat c$ \qquad The superVirasoro  level of an $\CN=2$ superconformal  algebra.

\medskip
$e(x) $ \qquad $\exp(2\pi i x)$.

\medskip
$G$ \qquad 3D Newton constant. Has dimensions of length.

\medskip
$\Gamma$ \qquad $SL(2,\IZ)$.

\medskip
$\Gamma_\infty$ \qquad The subgroup
of $\Gamma$ stabilizing $\tau = i\infty$.

\medskip
$j$ \qquad An $SU(2)$ spin. $j\in \half \IZ_+$. Associated with
harmonics of $S^3$.

\medskip
$J$ \qquad 2+1 black hole spin on $AdS_3$.

\medskip
$J(z)$ \qquad Leftmoving $U(1)$ current in a $d=2,\CN=2$
superconformal algebra.

\medskip
$\tilde J(z)$ \qquad Rightmoving  $U(1)$ current in a $d=2,\CN=2$
superconformal algebra.

\medskip
$J_0$ \qquad The zeromode of the leftmoving current $J(z)$. For
$\CN=4$ reps $J_0 = 2 J_0^3$ has integral eigenvalues.

\medskip
$\tilde J_0$ \qquad The zeromode of the rightmoving current
$\tilde J(z)$.

\medskip
$J_0^3$ \qquad  $\half$-integer moded. From the $d=2,\CN=4$
superconformal algebra.

\medskip
$k$ \qquad  The level $k=Q_1 Q_5$ in the D1 D5 system. Positive integral.

\medskip
$\ell$ \qquad  The radius of curvature of $AdS_3$.
 and
its quotients. $\Lambda = -1/\ell^2$.  In the $D1D5$ system $\ell^2 = g_6 k$.

\medskip
$\ell$ \qquad  Also, a nonnegative integral eigenvalue of $J_0$ in
the CFT with target ${\rm Sym}^kX$.

\medskip
$L_0, \tilde L_0$ \qquad Left, right Virasoro generators. These
are dimensionless.

\medskip
$q$ \qquad \qquad  $q= \exp[2 \pi i \tau] $.

\medskip
$Q_1, Q_5$ \qquad Positive integer numbers of $D1,D5$ branes.

\medskip
$\omega, \tilde\omega$ \qquad Fugacities for $SU(2)_L \times SU(2)_R$
spins. These are complex.

\medskip
$X$ \qquad A Calabi-Yau manifold.

\medskip
$y$ \qquad $y= e^{2\pi i z}$ in the context of Jacobi forms and
$y= e^{2\pi i \omega }$ in the context of black hole statistical
mechanics.

 \medskip
$\CZ_f, \CZ_\phi$ \qquad \Holographic transform of a
modular form $f$ or Jacobi form $\phi$.

\medskip
$Z_{NS}, Z_{R}$ \qquad Partition functions.

\appendix{B}{First proof of the Rademacher expansions}

\subsec{Preliminaries:
Ford circles,   Farey fractions,  and
Rademacher paths}

Before proving the above result we need to give a
few preliminary definitions and results.

\ifig\radpathi{The Rademacher path for $N=1$
goes from $A$ to $B$ to $C$. }
{\epsfxsize3.0in\epsfbox{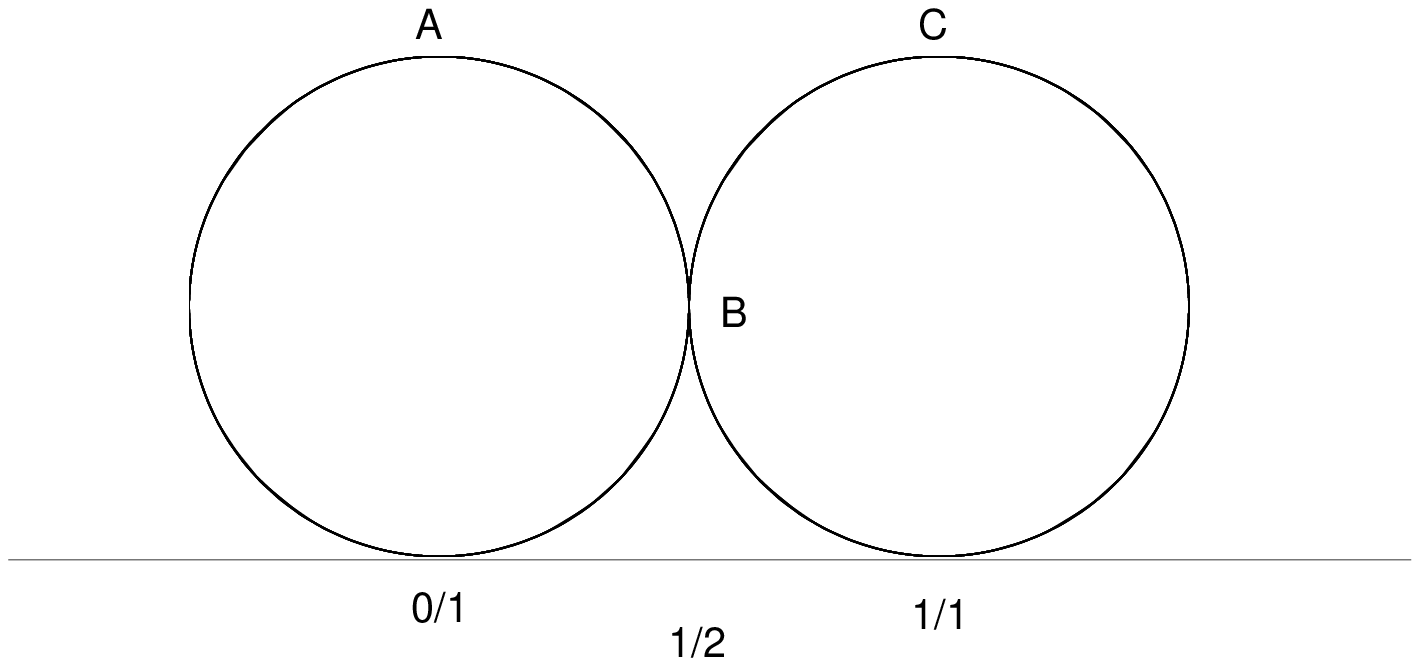}}

\ifig\radpathiii{Ford circles used to construct the Rademacher
path for $N=3$} {\epsfxsize3.0in\epsfbox{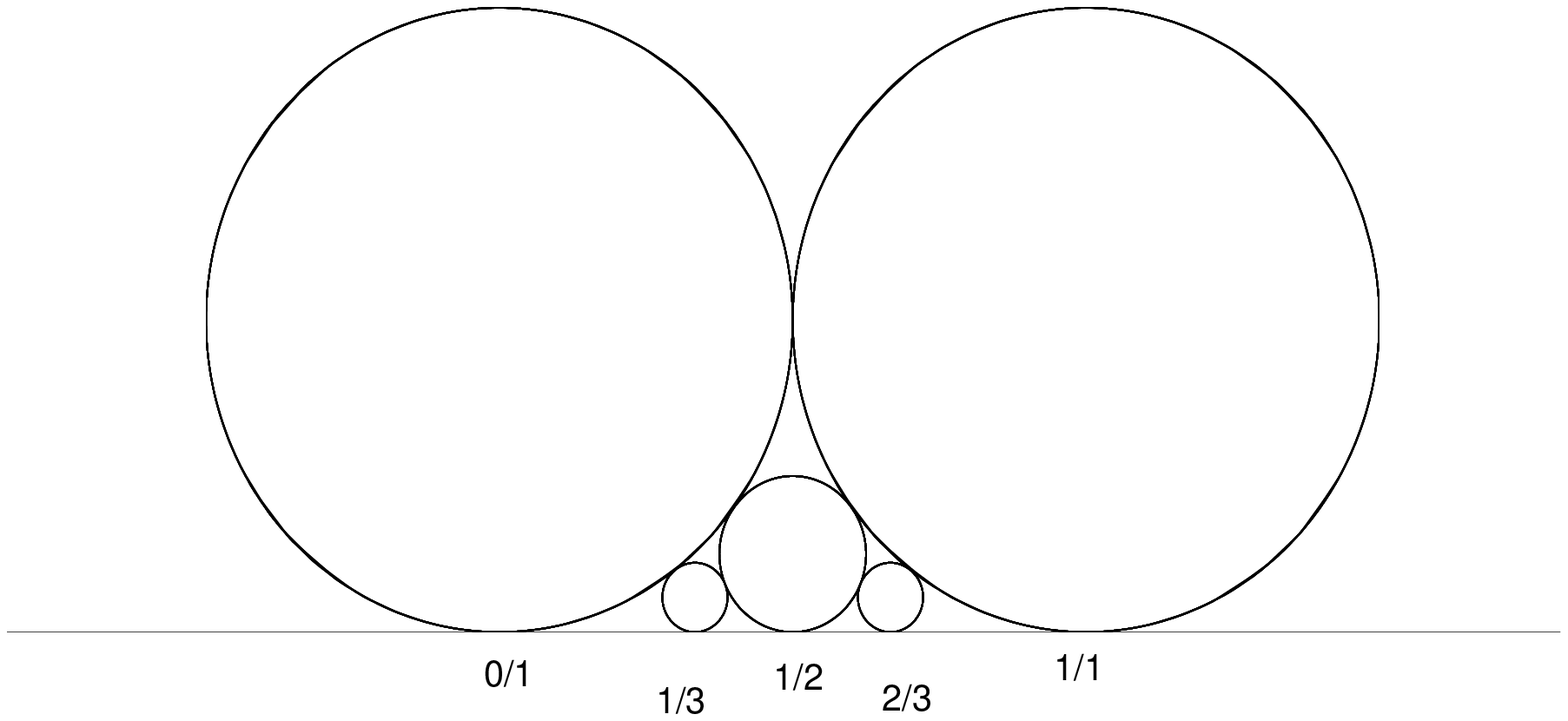}}

\ifig\radpathv{Ford circles used in the Rademacher
path for  $N=5$}
{\epsfxsize4.0in\epsfbox{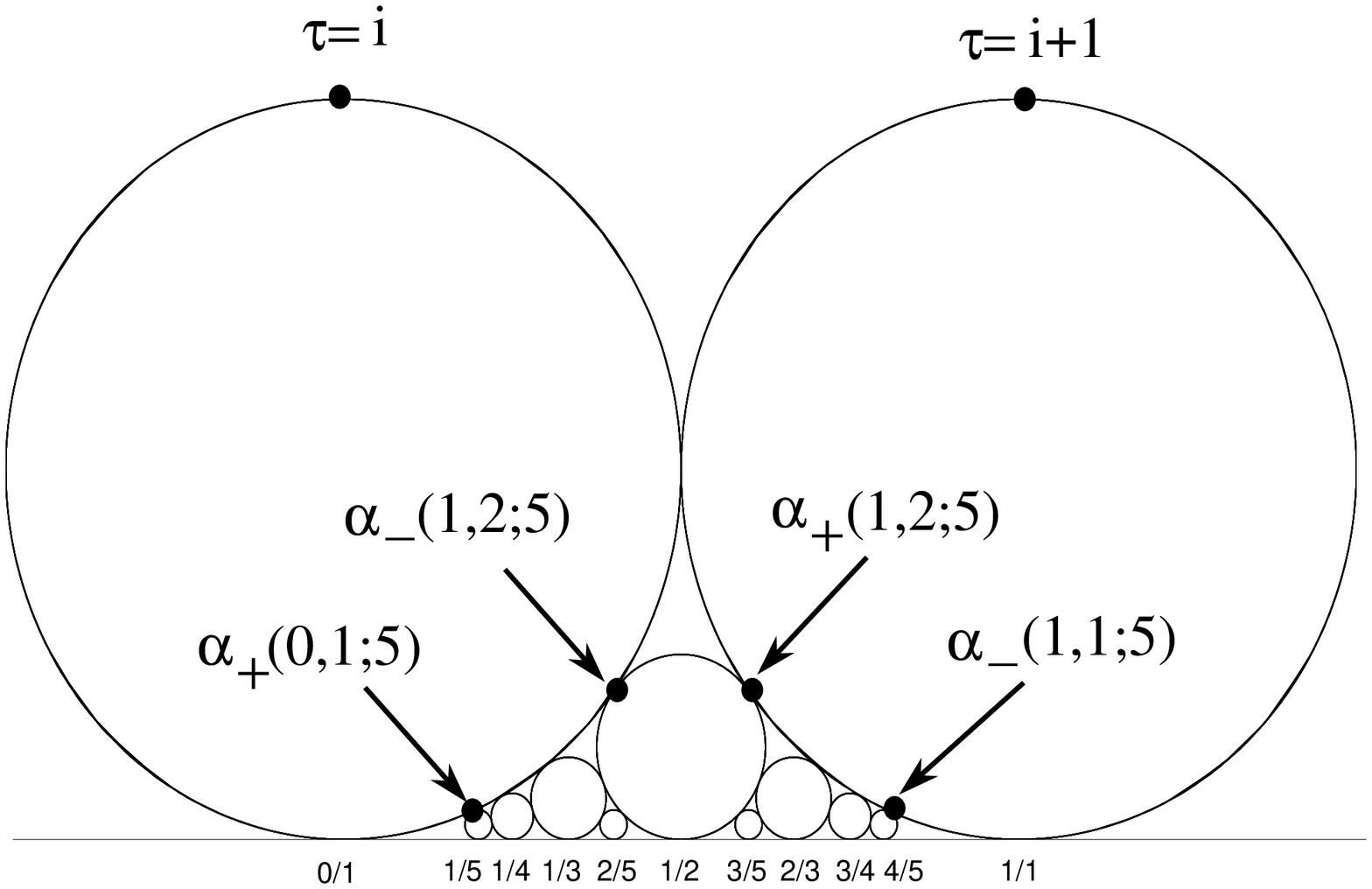}}

\ifig\zcirc{Path of integration for $z$. Let $\beta = 1/z$, then the
path is vertical, parallel to the imaginary $z$-axis. }
{\epsfxsize3.0in\epsfbox{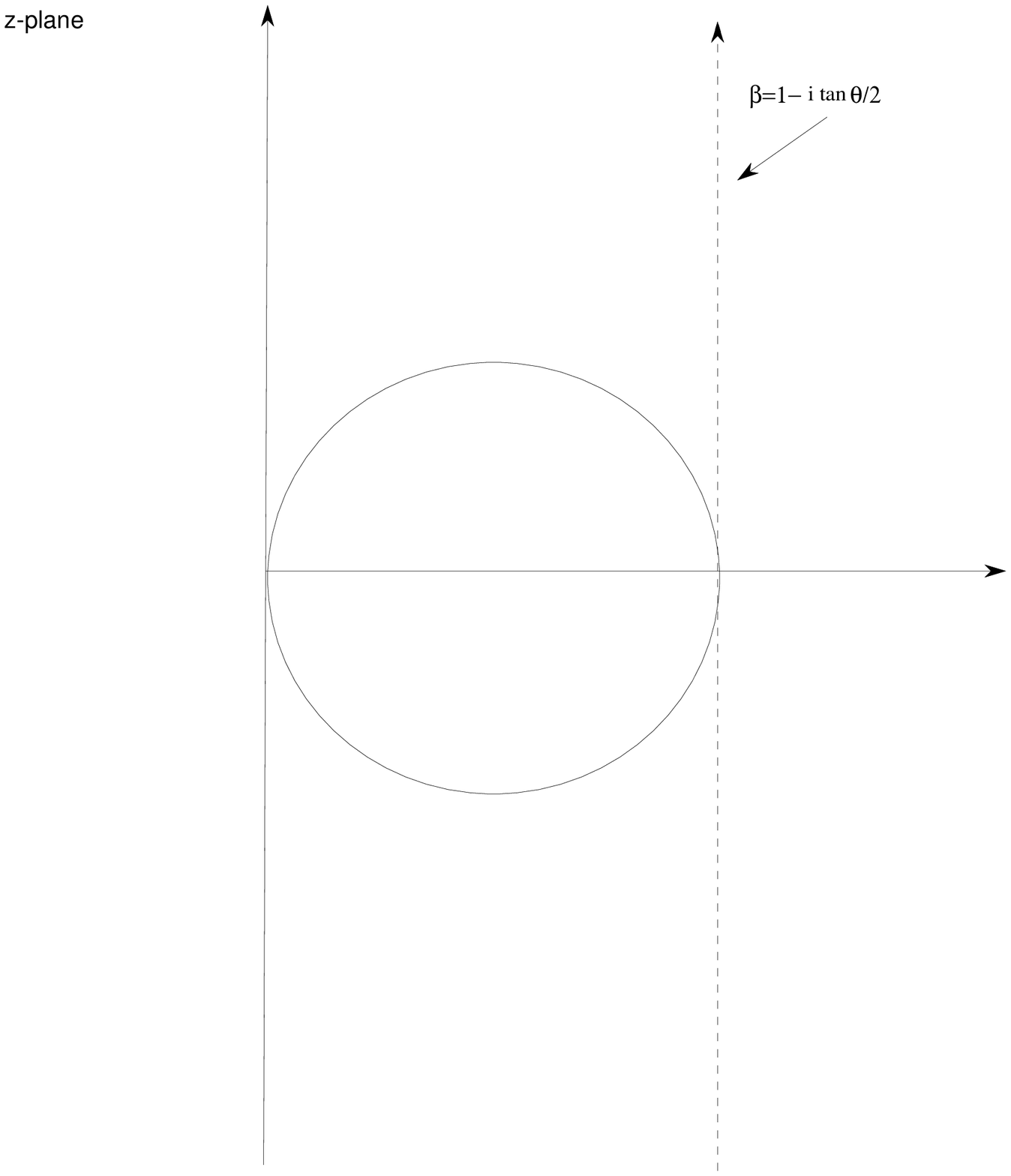}}

{\bf Definition:} The Ford circle $\CC(d,c)$ is defined for $d/c$
in lowest terms to be the circle of radius $1/(2 c^2)$ tangent to
the $x$-axis at $d/c$. It is given analytically by:
\eqn\frdcrcl{
\tau(\theta) := { d \over  c} + {i \over  c^2} z(\theta) = { d
\over c} + {i \over  c^2} ({1+ e^{i \theta}\over  2} ) }
Note that
$z$ runs over a circle of radius $1/2$ centered on $1/2$ shown in
\zcirc: \eqn\frdzee{ z(\theta):= {1+ e^{i \theta}\over  2}  =
\cos(\theta/2) e^{i \theta/2} }

{\bf Definition:}. The Farey numbers $\CF_N$ are the fractions in
lowest terms between $0$ and $1$ (inclusive) with denominator $\leq
N$.

We will now need some facts about Ford circles and Farey numbers.
These can all be found in \apostol, ch. 5, and in \rademacherii.
First, two Ford circles are always disjoint or tangent at exactly
one point.  Moreover, two Ford circles $\CC(d_1,c_1)$ and
$\CC(d,c)$ are tangent iff $d_1/c_1 < d/c$ are consecutive numbers
in some Farey series (\apostol, Theorem 5.6). If $d/c\in \CF_N$
then denote the neighboring entries in the Farey series $\CF_N$ by
$d_1/c_1 < d/c < d_2/c_2$. Call the intersection points of the
Ford circle $\CC(d,c)$ with the neighboring Ford circles
$\alpha_-(d,c;N), \alpha_+(d,c;N)$. Explicit formulae    for
$\alpha_\pm(d,c;N)$, are (  \apostol, Thm. 5.7): \eqn\alphplmn{
\eqalign{ \alpha_-(d,c;N) & = {d\over  c} - {c_1 \over
c(c^2+c_1^2)} + {i \over  c^2 + c_1^2} \cr \alpha_+(d,c;N) & =
{d\over  c} + {c_2 \over  c(c^2+c_2^2)} + {i \over  c^2 + c_2^2}
\cr} } In mapping to the $z$-plane by \frdcrcl\ we get
\eqn\zeeplmn{ \eqalign{ z_-(d,c;N)  & =   {c^2 \over  (c^2+c_1^2)}
+ {i cc_1 \over  c^2 + c_1^2} \cr z_+(d,c;N) & =   {c^2 \over
(c^2+c_2^2)} -  {i cc_2 \over  c^2 + c_2^2} \cr} } A key property
used in the estimates below is that if $z$ is on the chord joining
$z_-(d,c;N)$ to $z_+(d,c;N)$ then (  \apostol, Thm. 5.9):
\eqn\chordest{ \vert z \vert < {\sqrt{2} c \over  N} }

\bigskip
\noindent {\bf Definition:} The Rademacher path $\CP(N)$ is
\eqn\radpth{ \CP(N) = \cup_{d/c \in \CF_N} \gamma^{(N)}_{d,c} }
where $\gamma^{(N)}_{d,c}$ is the arc of $\CC(d,c)$ above $d/c$
which lies between the intersection with the Ford circles of the
adjacent Farey fractions in $\CF_N$. Call the intersection points
$\alpha_-(d,c;N), \alpha_+(d,c;N)$ as above. We orient the path
from $\alpha_-$ to $\alpha_+$.

Qualitatively, as $N \rightarrow \infty$ the Rademacher
path approaches more and more closely an integration
around the complete Ford circles associated with
all the rational numbers in $[0,1)$. The arcs associated
with $\CC(0,1), \CC(1,1)$ require special treatment,
since the integration path is only over a half-arc. However,
by translating $\CC(1,1)$ under $\tau \rightarrow \tau-1$
these two arcs become a good approximation to the
integral over the full Ford circle $\CC(0,1)$.

\subsubsec{Modular transformations of Ford circles}

We will make modular transformations on the Ford circles to what
we call the ``standard circle.'' This is the circle: \eqn\frdtrmn{
\tau(\theta) = iz(\theta)= i {1+ e^{i \theta}\over  2}  = i
\cos(\theta/2) e^{i \theta/2} . } Under the modular transformation
$ \tau \rightarrow -1/\tau$ the standard circle maps to a line
parallel to the $x$-axis:
 \eqn\tmstdrd{ -1/\tau(\theta) =
\tan(\theta/2) + i } It is also very useful to introduce the
parameter \eqn\betathre{ \beta(\theta) :=1/z(\theta)  = 1- i
\tan(\theta/2) }

We see from the above that the
 modular transformation $\tau \rightarrow -1/\tau$
takes the Ford circle $\CC(0,1)$ to the line
$\Im \tau=1$. In fact, any Ford circle $\CC(-d,c)$ can be mapped to the
line $\Im \tau=1$ by a transformation of the form
\eqn\frdtmii{
\gamma_{c,d} = \pmatrix{ a & b \cr c & d  \cr} \qquad ad - bc = 1
}
$\gamma_{c,d}$ is well-defined up to
left-multiplication by $\Gamma_\infty$.
See \mtrxmlt.

\subsec{Proof of the theorem}

We now give the proof of \radfirst\kernelfn\kernelfnii. Using
\collec\ we have \eqn\proofi{ F_\nu(n) = \int_\gamma d \tau e^{ -
2 \pi i (n+ \Delta_\nu)\tau} f_\nu(\tau), } which holds for any
path $\gamma$ in the upper half plane with $\gamma(1) =
\gamma(0)+1$. In particular, we may we take the Rademacher path
$\CP(N)$: \eqn\proofii{ F_\nu(n) =  \sum_{d/c\in \CF_N }
\int_{\gamma^{(N)}(d,c) } d \tau e^{ - 2 \pi i (n +
\Delta_\nu)\tau } f_\nu(\tau) }
Written out explicitly this is:
\eqn\proofii{ \eqalign{ F_\nu(n) =
\int_{\alpha_-(1,1;N)-1}^{\alpha_+(0,1;N)}  & d \tau e^{ - 2 \pi i
(n + \Delta_\nu)\tau } f_\nu(\tau) \cr + \sum_{d/c\in \CF_N, 0<
d/c < 1} \int_{\alpha_-(d,c;N)}^{\alpha_+(d,c;N)} & d \tau e^{ - 2
\pi i (n + \Delta_\nu)\tau } f_\nu(\tau) \cr}
}
 In the case $d/c = 0/1, 1/1$ we should translate
the arc above $\CC(1,1)$ by $\tau \rightarrow \tau-1$ to get a
single arc above $\CC(0,1)$. Denote the integrals in \proofii\
over the arc $\gamma^{(N)}(d,c)$ by $\CI_\nu(d,c;N)$.

Now   for each of the integrals in the Ford circles we make a
modular transformation of the form \frdtmii\ which maximizes the
imaginary part of the top of the Ford circle. \foot{Under modular
transformations the maximal imaginary part of the image of the top
of a Ford circle is achieved by the transformations \frdtmii: We
wish to maximize ${ \Im \tau \over  \vert c' \tau + d' \vert^2 }$
for $\tau = {d \over  c}  + {i \over  d^2} z $ for $z\cong 1$.
Clearly we can minimize the denominator by taking
$\gamma_{c',-d'}$ with  $c' = c, d'= d $. } This brings the Ford
circle to a standard circle, which we can take to be the
$z$-circle, centered on $z=1/2$, or the ``circle'' given by $i/z$
which is the  line $\Im \tau=1$: \eqn\proofiii{ \eqalign{ \tau & =
d/c + i z/c^2 \cr \tau' & = \gamma_{c,-d} \cdot \tau = a/c + i/z =
a/c + \tan(\theta/2) + i \cr}} So, using the modular
transformation law for $f_\mu$ the resulting integral for this arc
is
\eqn\proofiv{
\eqalign{ \CI_\nu(d,c;N) = & c^{w-2} e^{ - 2 \pi i (n +
\Delta_\nu)(d/c) } M^{-1}_{\nu\mu}(\gamma_{c,-d} )\cr & i
\int_{z_-(d,c;N)}^{z_+(d,c;N)} dz  z^{-w} e^{ 2 \pi  (n +
\Delta_\nu)z/c^2 } f_\mu(a/c + i/z )\cr} }
with the integral on $z$ over a circle of radius 1/2 shown in
\zcirc, with the orientation given by integrating from $\theta\cong
+\pi-\epsilon$ to $\theta \cong -\pi+\epsilon$. For $c>0$ and $w$
half-integral we use the principal branch of the logarithm.

We now split the Fourier sum for $f_\mu$ into its polar and
nonpolar pieces
\eqn\proofv{ \eqalign{ f_\mu(\tau) & = f^-_\mu(\tau) +
f^+_\mu(\tau) \cr f^-_\mu(\tau) & := \sum_{m+\Delta_\mu < 0}
F_\mu(m) e^{ 2 \pi i (m+ \Delta_\mu) \tau} \cr f^+_\mu(\tau) & :=
\sum_{m+\Delta_\mu \geq  0} F_\mu(m) e^{ 2 \pi i (m+ \Delta_\mu)
\tau} \cr} } and similarly define $\CI_\nu^\pm$ such that
$\CI_\nu(d,c;N)= \CI_\nu(d,c;N)^- + \CI_\nu(d,c;N)^+$.

The integral   $\CI_\nu(d,c;N)^-$ will become the sum of
$I$-Bessel functions for $N \rightarrow \infty$. We will show that
$\sum_{d/c \in \CF_N } \CI_\nu(d,c;N)^+$ goes to zero for $N
\rightarrow +\infty$.

\ifig\contourdef{Contours for estimates in the
Rademacher series. If $(m+\Delta_\mu)\geq 0$ we can
deform the contour in $\beta(\theta)$ into the right
half-plane and obtain zero in the $N \rightarrow \infty$
limit. }{\epsfxsize3.0in\epsfbox{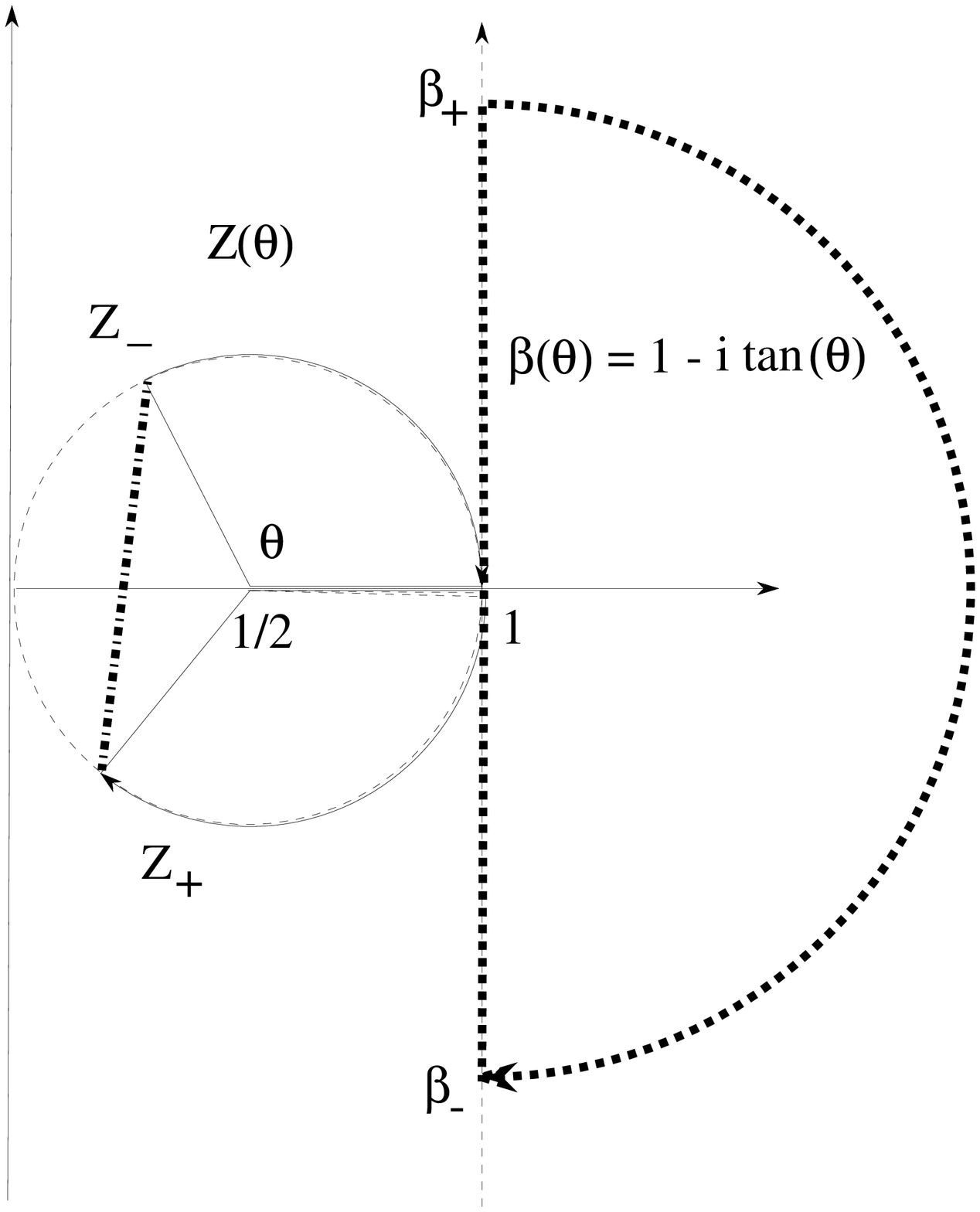}}

The basic integral we need to estimate is:
\eqn\bscintgrl{
\int_{z_-(d,c;N)}^{z_+(d,c;N)} dz z^{-w}
e^{ 2 \pi   (n + \Delta_\nu)z/c^2  }
e^{- 2 \pi (m+\Delta_\mu)/ z }
}
The integral is over the circular contour in
\contourdef, oriented from $z_-$ to $z_+$.
Already saddle point techniques show that
the behavior of the integral is very different
depending on the sign of $(m+\Delta_\mu)$.

  If $(m+\Delta_\mu) \geq 0 $ then we
deform in integral along the arc $z(\theta)$ to an integral along
the chord joining $z_-(d,c;N)$ to $z_+(d,c;N)$ shown in
\contourdef. More fundamentally, in terms of $\beta(\theta)$, we
can deform the contour into the right half-plane. As $N
\rightarrow \infty$, $z_\pm(d,c;N) \rightarrow 0$. Therefore, the
chord is near zero. Along the chord if we write $z= \epsilon e^{i
\phi} $ then $\Re(1/z) = (1/\epsilon)  \cos(\phi) $ can get very
large. Therefore this contour deformation will only be useful for
$(m+\Delta_\mu) \geq 0 $. In this case, the minimal value of
$\Re(1/z)$ is taken by   the points on the Ford circle, so for $z$
on the chord:
\eqn\proofvi{ \Re(1/z) \geq  Min[\Re(1/z_+),\Re(1/z_-) ] = 1 }
Therefore, a crude estimate of \bscintgrl\ is \eqn\bscintgrlii{
\eqalign{ \biggl\vert \int_{z_-(d,c;N)}^{z_+(d,c;N)} dz z^{-w} e^{
2 \pi   (n + \Delta_\nu)z/c^2  } e^{- 2 \pi (m+\Delta_\mu)/ z
}\biggr \vert & \leq \vert z_+(d,c;N) - z_-(d,c;N) \vert \cdot
 \cr
\cdot ({\rm Max}_{\rm chord} \vert z \vert^{-w} )\cdot
\bigl( {\rm Max}_{\rm chord} \vert e^{ 2 \pi/c^2   (n +
\Delta_\nu) z }  \vert \bigr) \cdot &
e^{- 2 \pi (m+\Delta_\mu) } \qquad (m+\Delta_\mu) \geq 0 \cr}
}
Using \chordest\ ( \apostol, Thm 5.9 ) we have
$\vert z \vert \leq \sqrt{2} c/N$ along the chord.
Therefore
\eqn\anothest{
  {\rm Max}_{\rm chord} \vert e^{ 2 \pi/c^2   (n +
\Delta_\nu) z }  \vert =
{\rm Max}_{\rm chord}  e^{ 2 \pi/c^2   (n +
\Delta_\nu) Re(z) }  \leq
{\rm Max} [1, e^{2 \pi (n + \Delta_\nu) \sqrt{2}/(N c) }]
}
To estimate the other factors in \bscintgrl\  we now take
  $w \leq 0$ and use \chordest\ so:
\eqn\bscintgrlii{
\eqalign{
\biggl\vert \int_{z_-(d,c;N)}^{z_+(d,c;N)} dz z^{-w}
e^{ 2 \pi/c^2   (n + \Delta_\nu)z  }
e^{- 2 \pi (m+\Delta_\mu)/ z }\biggr \vert
&
\leq \cr
2 \bigl({ \sqrt{2} c \over  N}\bigr)^{1-w} {\rm Max} [1, e^{2 \pi (n +
\Delta_\nu) \sqrt{2}/(N c) }]
&
 e^{- 2 \pi (m+\Delta_\mu) } \qquad (m+\Delta_\mu) \geq 0 \cr}
}

Using the absolute convergence of $F_\nu(\tau=i)$ we see that the
sum on $m$ in $f^+$ gives an $N$-independent constant and we need
to estimate:
\eqn\morests{ \eqalign{ \biggl \vert \sum_{d/c\in\CF_N}
\CI_\nu(d,c;N)^+ \biggr \vert & \leq \sum_\mu \sum_{d/c\in\CF_N }
\vert M^{-1}_{\nu\mu}(\gamma_{c,-d}) \vert c^{w-2}\cr & 2 \bigl({
\sqrt{2} c \over  N}\bigr)^{1-w}  e^{2 \pi \vert n +
\Delta_\nu\vert \sqrt{2}/(N c) } \vert f_\mu^+(\tau=i)\vert\cr} }
We can now estimate the error from \morests\ as
follows:

1. The sum  on $d$ has $\phi(c) \leq c$ terms.

2. $\vert M^{-1}_{\mu\nu}(\gamma_{c,d}) \vert \leq 1$
because $M_{\mu\nu}$ are unitary matrices

3. The sum on $c$ has $  N$ terms.

Thus  we get an upper bound of
\eqn\upperest{ \biggl \vert \sum_{d/c\in\CF_N} \CI_\nu(d,c;N)^+
\biggr \vert \leq N^{w}~ r ~ 2^{(3-w)/2} {\rm max}_{\mu} \vert
f^+_\mu(\tau=i) \vert
 e^{2 \pi \vert n + \Delta_\nu\vert \sqrt{2}/N} }
For $w<0$ this goes to zero for $N \rightarrow \infty$.

  Now we come to the finite
number of terms with  $(m+\Delta_\mu) < 0 $. In \bscintgrl\ we
write
 \eqn\bscintgrl{ \eqalign{ \int_{z_-(d,c;N)}^{z_+(d,c;N)} dz
z^{-w} e^{ 2 \pi/c^2   (n + \Delta_\nu)z  } e^{- 2 \pi
(m+\Delta_\mu)/ z } & = \qquad\qquad\qquad\cr
\int_{0}^{z_+(d,c;N)} dz z^{-w} & e^{ 2 \pi/c^2   (n +
\Delta_\nu)z  } e^{- 2 \pi (m+\Delta_\mu)/ z }  \cr +
\int_{z_-(d,c;N)}^{0} dz z^{-w} e^{ 2 \pi/c^2   (n + \Delta_\nu)z
} e^{- 2 \pi (m+\Delta_\mu)/ z } & + \oint  dz z^{-w} e^{ 2
\pi/c^2   (n + \Delta_\nu)z  } e^{- 2 \pi (m+\Delta_\mu)/ z } \cr}
} The last integral is essentially the Bessel function of the
theorem. All integrals are along the Ford circle. However, along
the circle $\Re(1/z) = \Re(1 - i \tan(\theta/2) )= 1$, while $$
\Re(z) \leq \Re(z_+) \leq \vert z_+ \vert = {c \over  \sqrt{c^2 +
c_2^2} } $$ so once again we use \chordest\ to get the estimate on
the error \eqn\bscintgrliiii{ \eqalign{ \biggl\vert
\int_{0}^{z_+(d,c;N)} dz z^{-w} e^{ 2 \pi/c^2   (n + \Delta_\nu)z
} e^{- 2 \pi (m+\Delta_\mu)/ z }  \biggr\vert & \leq   \pi \vert
z_+(d,c;N)   \vert \cdot
 \cr
\cdot ({\rm Max}_{\rm arc} \vert z \vert^{-w} )\cdot
\bigl( {\rm Max} [1, e^{2 \pi (n + \Delta_\nu) \sqrt{2}/(N c) }]
 \bigr) \cdot  &
e^{- 2 \pi (m+\Delta_\mu) } \cr
 \leq \pi \bigl({ \sqrt{2} c \over  N}\bigr)^{1-w}
& e^{2 \pi  \vert n + \Delta_\nu \vert \sqrt{2}/(N c) } e^{-
2\pi(m+\Delta_\mu)} \cr} } So again, as in \upperest,  the error
in dropping these terms is $\sim N^w$. Similar remarks apply to
the integral from $0$ to $z_-$.  This leaves the integral
\eqn\finintgl{
 \oint  dz z^{-w}
e^{ 2 \pi/c^2   (n + \Delta_\nu)z  }
e^{- 2 \pi (m+\Delta_\mu)/ z }
= - \int_{1-i\infty}^{1+i \infty} d \beta \beta^{w-2}
\exp\biggl[ {2 \pi \over  c^2} {n+ \Delta_\nu \over  \beta}
+ 2\pi \vert m+ \Delta_\mu \vert \beta \biggr]
}
which can be expressed in terms of the $I$-Bessel
function. Note
\eqn\intrbeta{
\beta = {1 \over  z} = 1 - i \tan(\theta/2)
}

The above argument can be extended to $w=0$.
See \knopp.

\appendix{C}{An elementary proof of the Rademacher
expansion}

In this appendix we give a much simpler
proof of the Rademacher expansion, making
use of the mathematical transformation
that appears when relating the conformal
field theory and supergravity partition functions.

For simplicity we take the case of a one-dimensional representation
of $SL(2,\IZ)$ without multiplier system and with integral weight
$w<0$.
 The proof is simply the following:

1. Observe that  $(q {\p \over  \p q})^{1-w} f$
transforms with modular weight $2-w>2$.

2. Note that  $(q {\p \over  \p q})^{1-w} f$
has no constant term and is
orthogonal to all the cusp  forms in $M_{2-w}$.
(We will prove this below.)

3.  Therefore
$F_\mu  := (q {\p \over  \p q})^{1-w} f_\mu$
is fully determined by applying the Poincar\'e
series operation to the polar part (negative powers
of $q$) $F^-_\mu$, since the difference of
two such would be a Poincare series defining
an ordinary modular form in $M_{2-w}$, that is,
a cusp form.
The Poincar\'e series will converge absolutely
for $w<0$.
Note that
\eqn\polarpart{
F_\mu^- = (q {\p \over  \p q})^{1-w} f_\mu^-
}

Now that we have weight $2-w>2$
and can represent $F_\mu$ as an Poincar\'e
series
we can  apply the Petersson formula
for Fourier coefficients of Poincar\'e series.
This gives exactly the Rademacher formula.

Proof of step 2: We will show that  the Petersson inner product is
\eqn\pip{
\int_{\CF} {dx dy \over  y^2} y^{2-w} (q {\p \over  \p q})^{1-w} f(\tau)
\bar g(\bar \tau) = 0 }
for cusp forms $g\in M_{2-w}$.
Since $f$ has a polar part the integral is understood in
the usual sense of cutting off $\Im\tau < \Lambda$ and
then taking $\Lambda\rightarrow \infty$.
 We can justify this using integration by parts with an operator
similar to \ohopiii, namely, $\nabla_W =( {\p\over \p \tau} + {W-2\over 2i y} ) $
which takes modular forms of weight $W-2$ to forms of weight $W$.
We have:
\eqn\ibp{
\int_{\CF} {dx dy \over y^2} y^W (\nabla_{W-2} f(\tau))\bar g(\bar \tau)=0
}
simply by integration by parts. Note we need
  $f(\tau)\bar g(\bar \tau)$ to have no constant term.
Now, for $w<0$ we have
%
\eqn\prodofnabs{
\biggl(2\pi i q {\p \over  \p q}\biggr)^{1-w} f(\tau) =  \nabla_{2-w} \nabla_{-w} \cdots
\nabla_{w+2} \nabla_{w} f(\tau)
}
and now step 2 follows.
$\spadesuit$

 The point of this derivation is that the proof
of Petersson's formula (we recall it below) is   more
elementary and straightforward
than Rademacher's method based on Farey series and Ford
circles. Note also that it is consistent with Lemma 9.1 of
\borcherds.

\subsec{Petersson's formula}

Here we recall a standard formula from analytic number theory.
See, for examples, texts by  Iwaniec  \iwaniec\ or Sarnak \sarnak\
for further discussion. We let $w>0$ be a positive weight.
Consider the Poincar\'e series:
\eqn\ps{ f_p(\tau) :=
\sum_{\Gamma_\infty \backslash \Gamma} (c \tau + d)^{-w}
p(\gamma\cdot \tau) }
This is well-defined if $p(\tau+1 ) =
p(\tau)$. Unless  $w$ is even integral we must fix $c>0$.
We will specialize to $p(\tau) = e^{2 \pi i m \tau} $, $m\in \IZ$.
If $w>2$ the series is absolutely convergent. Then $f_p(\tau)$ is
a modular form, although for $m<0$ we allow poles at the cusps.
(This is usually excluded, e.g. in \iwaniec\  one takes $m\geq 0$
but we want $m<0$ for our application. The sign of $m$ does not
change the convergence properties. )

Petersson's formula gives an expression for
the Fourier coefficients in
\eqn\petersson{
f_p(\tau) = \sum_{\ell\in \IZ}   F(\ell) q^\ell
}

The derivation is  elementary. We write
 \eqn\drvp{ \eqalign{
f_p(\tau) & = p(\tau) +  \sum_{( \Gamma_\infty \backslash
\Gamma/\Gamma_\infty)'} \sum_{\ell\in \IZ} \bigl[ c (\tau + \ell)+
d\bigr]^{-w} p({a (\tau + \ell) + b \over  c (\tau + \ell) + d})
\cr & = p(\tau) +  \sum_{( \Gamma_\infty \backslash
\Gamma/\Gamma_\infty)'} \sum_{\hat\ell\in \IZ}
\int_{-\infty}^{+\infty}  e^{-2 \pi i \hat \ell t} \bigl[ c (\tau
+ t)+ d\bigr]^{-w} p({a (\tau + t) + b \over  c (\tau + t) + d})
dt \cr} }
where we used the Poisson summation formula. Now we specialize
again to $p(\tau) = e(m \tau)$. By a simple change of variables
the integral in \drvp\ becomes \eqn\doint{ e(\hat \ell \tau + \hat
\ell d/c + m a/c) c^{-1} \int_{-\infty + i cy}^{+\infty + i cy}
e(-\hat \ell v/c - m/(cv) ) v^{-w} dv } where we use the standard
notation $e(x):= \exp[2 \pi i x]$. The contour integral does not
depend on $y$ by Cauchy's theorem, and for $\hat \ell\leq 0$ we
can close the contour in the upper halfplane and get zero. (For
$\hat\ell =0$ we must have $w>1$ for this.) For $\hat\ell>0$ we
close in the lower half-plane and the countour becomes a Hankel
contour surrounding the lower imaginary axis. This gives the
standard result in terms of the Bessel function $J_{w-1}$ for
$m>0, \hat \ell>0$:

\eqn\peterson{
  F(\ell) = -2 \pi i^{-w}
\sum_{c=1}^\infty {1 \over  c} Kl(\ell,m;c)
({\ell \over  m})^{(w-1)/2} J_{w-1}({4 \pi \over  c} \sqrt{m\ell})
}
where $Kl(\ell,m;c)$ is the Kloosterman sum:
\eqn\kloosterman{
Kl(\ell,m;c):=\sum_{d \in (\IZ/c\IZ)^*} e(\ell d/c) e(m d^{-1}/c)
}

We can also do the contour integral if $m<0, \ell>0$. In this case
we get an $I$-Bessel function. One quick way to see this is to use
the relation between J- and I- Bessel functions. (See, e.g.,
Gradshteyn and Ryzhik,  GR 8.406): $$ J_\nu(e^{i \pi /2} z) = e^{i
{\pi \over 2} \nu} I_\nu(z). $$

Doing the integral we get (for $p(\tau) = e(m \tau), m<0$):
\eqn\pforms{
    F(\ell) =   \sum_{c=1}^{\infty} {2 \pi \over  c} Kl(\ell, m;c)
\bigl({\ell \over  \vert m\vert }\bigr)^{(w-1)/2}   I_{w-1}\bigl({4 \pi \over  c}
\sqrt{\vert m\vert \ell} \bigr). }

\listrefs

\bye